**Strong selective sweeps associated with ampliconic regions in great ape X chromosomes**

Kiwoong Nam[1*], Kasper Munch[1], Asger Hobolth[1], Julien Y. Dutheil[2], Krishna Veeramah[3], August Woerner[3], Michael F. Hammer[3], Great Ape Genome Diversity Project, Thomas Mailund[1], Mikkel H. Schierup[1,4*]

[1]Aarhus University, Bioinformatics Research Centre, Aarhus, 8000, Denmark, [2]Institut des Sciences de l'Évolution, Université Montpellier 2, Montpellier, 34095, France, [3]University of Arizona, Arizona Research Laboratories, Tucson, AZ, 85721, USA, [4]Aarhus University, Department of Bioscience, Aarhus, 8000, Denmark

* Correspondence: kiwoong@birc.au.dk, mheide@birc.au.dk

**Abstract**

The unique inheritance pattern of X chromosomes makes them preferential targets of adaptive evolution. We here investigate natural selection on the X chromosome in all species of great apes. We find that diversity is more strongly reduced around genes on the X compared with autosomes, and that a higher proportion of substitutions results from positive selection. Strikingly, the X exhibits several megabase long regions where diversity is reduced more than five fold. These regions overlap significantly among species, and have a higher singleton proportion, population differentiation, and nonsynonymous to synonymous substitution ratio. We rule out background selection and soft selective sweeps as explanations for these observations, and conclude that several strong selective sweeps have occurred independently in similar regions in several species. Since these regions are strongly associated with ampliconic sequences we propose that intra-genomic conflict between the X and the Y chromosomes is a major driver of X chromosome evolution.

**Introduction**

Evolution is expected to progress faster in the X chromosome than in autosomes for a wide variety of reasons[1-3]. Genetic drift is stronger on the X chromosome since only three X chromosomes exist for each four autosomal chromosomes. Fully or partly recessive variants on the X chromosomes are more frequently exposed to selection than comparable variants on the autosomes because of the hemizygosity in males. Finally, the gene content of the human X chromosome possesses an enrichment of genes involved in reproduction, brain development and muscle development[4].

Several simple demographic processes affect the X chromosome differently, complicating the study of natural selection. First, population size changes will affect X chromosome diversity differently because X chromosome diversity equilibrates more rapidly to the new population size than autosomes[5]. Second, differentiation of X chromosomes depends on preferential migration of males or females[6]. Third, the mutation rate on the X chromosome is lower than that on the autosomes, due to a higher number of cell division in the male germ line where fewer X chromosomes reside[7,8]. Finally, differences in female and male effective population size due to particular mating patterns will affect relative diversity on the X and autosomes[9,10].

These potentially confounding factors contribute to the lack of consensus in the debate on whether or not the human X chromosome has been preferentially targeted by adaptive evolution[11]. Empirical evidence supports stronger signatures of selection on X chromosomes[12-15] as well as a stronger effect of demographic history on the diversity of human X chromosomes than on autosomes[12]. A recent study suggests hard selective sweeps to be rare in recent human evolution on both autosomes and X chromosomes[14], but exome sequencing[16] suggests that 30-40% of X-linked amino acid changes in the central chimpanzee lineage has been fixed by positive selection.

Here we present a comparative analysis of X chromosome evolution based on recently published data from the great ape genome diversity project[17]. The data allows independent investigation of the outcomes of recent X chromosome evolution in at least the four completely independent lineages of orang-utans, gorillas, chimpanzees and bonobos. At the same time the data permits analysis of adaptive evolution on all branches of the great apes phylogeny. We find that adaptive evolution on X chromosomes is much more prominent than on autosomes in most of the great ape species and that the X chromosome has recently been targeted by a large number of selective sweeps with very high selection coefficients.

**Results**

**Selection affects the X chromosomes differently than the autosomes**

The comparative analysis is based on 1.7 Gb of each genome of the nine (sub)species investigated, called from the mapping against human reference genome (Supplementary Table 1-3). Genetic diversity of each species is based on 4-27 mainly female individuals except for the eastern lowland gorilla with two males and one female. In all species, the diversity within the X chromosome is less than 75% of the autosomal diversity. Particularly low ratios are found in eastern lowland gorilla and bornean orang-utans, which both experienced a recent population bottleneck, an event expected to have a more immediate effect on the X chromosome[5,17].

In all species, exons show lower diversity than introns and intergenic sequences, and this pattern is more pronounced on the X chromosomes (Figure 1a). In all species, the diversity increases with distance from genes, but X chromosomes have a steeper relationship as seen from the positive correlations between distance and the diversity ratio of X chromosomes to autosomes[17] (Figure 1b bottom row). The relative reduction of diversity on both autosomal and X-linked exons increases with intergenic diversity when comparing among species (Supplementary Figure 1), suggesting a key effect of the effective population size for both negative and positive selection. In line with this, the proportion of nonsynonymous to synonymous polymorphisms decreases with intergenic diversity for both autosomes and X chromosomes when comparing among species (Supplementary Figure 2).

**Many strong selective sweeps on X chromosomes**

Figure 2 shows the patterns of diversity along the X chromosomes for all species, expressed in terms of the nucleotide diversity and the proportion of singleton variants in non-overlapping windows of size 100 kb (autosomal results are shown in Supplementary Figure 3). Strikingly, regions of several megabases (up to 15 Mb long) show much reduced diversity, a pattern that is not mirrored in autosomes. Above each panel in Figure 2 red bars represent regions where diversity is less than 20% of the X chromosomal average in each species (See Supplementary Figure 4 for the distribution of pi values). In many cases, reduced diversity is accompanied by an increased proportion of singletons among polymorphic sites measured in the same windows. Regions of reduced diversity overlap to a large extent among species, but most have normal levels of diversity in at least one species. The western lowland gorilla shows the most striking patterns, followed by the orang-utans, the western chimpanzee, and the other chimpanzees, whereas bonobos show less and more local reductions in diversity. The diversity pattern in eastern lowland gorilla is unresolved due to extreme overall reduction of diversity, possibly due to the small number of sampled X chromosomes, and therefore this species is omitted in subsequent analyses. Figure 3a shows the percentage of the 100 kb windows with average nucleotide diversity less than 20 % of the chromosomal average for autosomes and X chromosomes for each species along the diagonal. The off-diagonal numbers show the extent to which these regions overlap. Evidently, such reduced diversity regions are generally rare on the autosomes but very common on the X chromosome in several species and the overlap between genera is appreciable.

To search for possible explanations for the strong reductions in diversity we first investigated divergence patterns for evidence of a reduced mutation rate. Only a minor reduction compatible with reduced polymorphism in the ancestral species is indicated (Supplemental Figure 5), ruling out appreciable mutation rate variation. We know of no demographic scenario that could reduce diversity locally along X chromosomes but not on autosomes, leaving natural selection targeting the X chromosome as the only possible explanation. Since at most 10% of nucleotides in the regions and likely less is under evolutionary constraint[18], the reductions in diversity must result from indirect selection in the form of background selection or selective sweeps. From deterministic calculations we conclude that background selection is unlikely to reduce diversity to less than 70% over Mega base regions (see Supplement Information for details), leaving soft or hard selective sweeps as the only possible explanation for our observations. Computer simulations of models of soft and hard sweeps, summarized in Table 1 (see Supplementary Information for details), show that even with large selection coefficients (up to $s = 0.5$) and a low allele frequency at the onset of positive selection ($p_0 = 0.01$), soft sweeps are expected to strongly affect diversity only in regions less than 250 kb wide. Thus, a very large number of soft selective sweeps within the last 200 kya are required to explain the reduced diversity, and several independent soft sweeps are needed to reduce diversity to 20% within a window of just one mega base. Fewer hard selective sweeps would be required, but with the effective population sizes in the range of 10,000-50,000, even a selective coefficient of 0.1 is expected only to reduce diversity to 75% and 25% in regions of 1 Mb and 190 kb, respectively (Table 1). To explain >5 Mb-wide regions of low diversity from the expected effect of single hard sweeps, selection coefficients of 0.5 are required, otherwise several sweeps are needed in each of the large regions.

Hard selective sweeps are expected to have strong effects on allelic differentiation among populations, and the proportion of singletons among SNPs should increase in regions around each positively selected site. The absolute divergence between the two orang-utan species and among chimpanzee subspecies is indeed much higher in regions where one or both of the populations has reduced diversity (Figure 3b). The same patterns are evident for Fst among the chimpanzee sub-species and between the two orang-utans (see Supplementary Figure 6). In addition, the low diversity regions have a higher non-synonymous to synonymous substitution ratio than the rest of the X chromosome in all species by 30 - 50%, except for gorilla where the ratio is comparable (Supplementary table 4). This again supports that reduced diversity is derived from positive selection, rather than background selection.

**Positive selection on the X chromosome is prevalent among the great apes**
To quantify the overall amount of positive selection on the X chromosomes we estimated the alpha value, the proportion of positively selected sites among substitution, using McDonald-Kreitman inspired analyses[11,19] for each species separately. Divergence was measured to the ancestral great ape[17] for African great apes and to human for orang-utans. In addition to the traditional contrast between evolution of synonymous and non-synonymous sites within and between species, we also investigated promoters and 3'UTRs for evidence of positive selection by contrasting these with repeats in the same regions. For all species, the alpha values are much higher on the X chromosomes than on the autosomes, often exceeding 50% of all fixations (Figure 4a). For autosomes, the corresponding alpha values are often estimated as negative, suggesting segregation of slightly deleterious non-synonymous variations. The ratio of non-synonymous to synonymous polymorphisms (and corresponding values for promoter and 3'UTR) is lower in general on the X chromosomes than on the autosomes arguing against weaker purifying selection on X chromosome (Figure 4b). Furthermore, enrichment of non-synonymous singletons on the autosomes is seen in all species, while on the X chromosome only three out of nine species have the enrichment of singletons for the non-synonymous SNPs (figure 4c). Taken together, these results suggest that the alpha values for the X chromosomes are not inflated by an increased fixation of slightly deleterious non-synonymous substitutions.

**Discussion**

The availability of many genomes from all the great ape species and subspecies allows us to observe independent realizations of the evolutionary processes shaping genetic diversity on the X chromosome. Several theoretical studies have suggested that adaptive evolution is expected to be more prominent on the X chromosome[20,21] and this has been supported by empirical observations in chimpanzees[16].

A number of phenomena may explain the prevalence of positive selection on the X chromosome. First, recessive beneficial mutations on X chromosomes are immediately selected upon in the heterozygous males, whereas such mutations on autosomes must reach appreciable frequencies before selection becomes effective. Thus, if new beneficial mutations are generally recessive, mutations on the X chromosome are more likely to be fixed[16,20]. Second, sexually antagonistic selection is also expected to be more frequent on the X chromosome. Recessive mutants beneficial to males are more likely to fix because they are subject to stronger selection even when deleterious effect on females far exceeds the beneficial effect on males[21,22]. Third, male reproduction genes are enriched on the X chromosome[23]. Such genes are typically under strong selective pressures[24,25] and may contribute to adaptive evolution of the X chromosome. Whereas these explanations may account for the large proportion of sites fixed by positive selection as measured by the McDonald Kreitman test they do not easily explain how positive selection produce the surprising abundance of mega base wide depressions in diversity in gorillas, chimpanzees and orang-utans.

While it is possible that selection on single new mutations (hard sweeps) may have produced each trough in diversity this would require selection coefficients at least twice as large as those acting on the lactase gene[26], the strongest known selective sweep in humans. Alternatively, numerous sweeps may have acted together to produce these regions. This is plausible considering the overlap of regions with the depressed diversity among species. To investigate possible causes of such repeated strong selection we performed gene ontology analysis to test if genes in a specific functional category are enriched in these regions. We found that genes associated with 'nuclear RNA export factor complex' are significantly enriched (corrected p value = 0.0287). These genes include *NXF3* and *NXF2*, which are involved in spermatogenesis[27,28]. Genes expressed exclusively in human testis are not enriched in these regions ($p = 0.260$, Supplementary Table 5), but further gene expression data from non-human great apes is necessary to support this conclusion.

A recent study[29] found that both human and mouse X chromosomes possess many ampliconic regions expressed exclusively in testis, and in mouse, 273 ampliconic genes are also expressed in postmeiotic cells[30]. These regions have a very dynamic turnover process of genes compared to regions with singly-copy regions, which are well preserved between human and mouse in support of Ohno's law[29]. Overlaying the diversity patterns with the positions of ampliconic regions we observe a striking concordance with regions of putative selective sweeps in all species (green lines in Figure 2). The enrichment of ampliconic regions is highly significant (Table 2), also when correcting for variation in gene density along the chromosome (Supplementary Table 6).

We hypothesize that the X-linked ampliconic regions are associated with selection for germ cell expression and intra-genomic conflict with genes on the Y-chromosome. A new duplication in an ampliconic region may be strongly selected for if it leads to preferential transmission of the X or Y chromosome, either during spermatogenesis or through competition between sperm cells carrying different sex chromosomes. A segregation distortion of >10% is more easily envisioned than similar selection coefficients acting on organism fitness alone. In response to fixation of a segregation distorter, compensatory mutations will be under very strong selection as well. Strong selective sweeps on both the X and Y chromosomes may be the result of such an arms race. If true, this suggests that intra-genomic conflict and co-evolution between ampliconic genes on X and Y chromosomes is a major driver of sex chromosome evolution in the great apes. In line with this conjecture, correlated gene amplification in mouse between sexually antagonistic X-linked *Slx* genes and Y-linked *Sly* genes, has been suggested to be a consequence of intra-genomic conflict between X and Y chromosomes balancing the sex ratio[31,32].

Intra-genomic conflict of the X and Y as outlined here may also explain the specific role for the X chromosome in hybrid depression among recently formed species and provide a major role in speciation among the primates. This is in line with a recent study showing that Neanderthal introgression in extant humans is severely depressed on the X chromosome[33]. Improved assembly of ampliconic regions in all great ape species, investigation of standing copy number variation of the regions and sequencing of the Y chromosomes will be pivotal for testing this hypothesis.

**Methods Summary**

We used the pipelines of the whole genome sequences and the variants of the great apes species from the great ape genome diversity consortium[17], mapped against hg18, followed by filtering out pseudo-autosomal regions, uncalled positions in any species, and heterozygous positions in male X chromosomes. The sequences were annotated with the refSeq and the ANNOVAR software[34]. The promoter is defined as the DNase I hypersensitive sites[35] located within 5 kb upstream regions of the transcription start sites and in the 5' UTR sequences. The pi value, the nucleotide diversity, was calculated using the formula of Nei and Li[36]. To calculate the pi values according to the distance from genes, we grouped all positions according to the distance from the nearest transcript by 5 kb, followed by calculating the pi value in each group. For each non-overlapping 100 kb window across the genomes, the pi value and the proportion of singletons among SNPs were calculated from called sites. All windows with less than 30% of called sequence were excluded. The population differentiation was estimated from mean allele difference of all SNPs and Fst values in the same windows[37,38]. Statistical analysis to find the enrichment of reproduction and ampliconic genes was performed using the fisher's exact test. The gene ontology test was performed using the GOrilla software[39]. For calculating the alpha value, we used the site frequency spectrum to control for the effect from slightly segregating deleterious mutations using the DFE alpha[11] and used the pre-calculates of ratio of nonsynonymous to synonymous sites in the amniotes[40].

**References**

1   Ellegren, H. The different levels of genetic diversity in sex chromosomes and autosomes. *Trends Genet.* **25**, 278–284 (2009).
2   Meisel, R. P. & Connallon, T. The faster-X effect: integrating theory and data. *Trends Genet.* **29**, 537-544 (2013)
3   Vicoso, B. & Charlesworth, B. Evolution on the X chromosome: unusual patterns and processes. *Nat. Rev. Genet.* **7**, 645–653 (2006)
4   Vallender, E. J. & Lahn, B. T. How mammalian sex chromosomes acquired their peculiar gene content. *Bioessays* **26**, 159–169 (2004).
5   Pool, J. E. & Nielsen, R. Population size changes reshape genomic patterns of diversity. *Evolution* **61**, 3001–3006 (2007).
6.  Laporte, V. & Charlesworth, B. Effective population size and population subdivision in demographically structured populations. *Genetics* **162**, 501–519 (2002).

7       Ellegren, H. Characteristics, causes and evolutionary consequences of male-biased mutation. *Proc. R. Soc. B* **274**, 1–10 (2007).
8       Makova, K. D. & Li, W.-H. Strong male-driven evolution of DNA sequences in humans and apes. *Nature* **416**, 624–626 (2002).
9       Hammer, M. F., Mendez, F. L., Cox, M. P., Woerner, A. E. & Wall, J. D. Sex-biased evolutionary forces shape genomic patterns of human diversity. *PLoS Genet.* **4**, e1000202 (2008).
10.     Evans, B. J. & Charlesworth, B. The effect of nonindependent mate pairing on the effective population size. *Genetics* **193**, 545–556 (2013).
11      Eyre-Walker, A. & Keightley, P. D. Estimating the rate of adaptive molecular evolution in the presence of slightly deleterious mutations and population size change. *Mol. Biol. Evol.* **26**, 2097–2108 (2009).
12      Gottipati, S., Arbiza, L., Siepel, A., Clark, A. G. & Keinan, A. Analyses of X-linked and autosomal genetic variation in population-scale whole genome sequencing. *Nat. Genet.* **43**, 741–743 (2011).
13      Hammer, M. F. et al. The ratio of human X chromosome to autosome diversity is positively correlated with genetic distance from genes. *Nat. Genet.* **42**, 830–831 (2010).
14      Hernandez, R. D. et al. Classic selective sweeps were rare in recent human evolution. *Science* **331**, 920–924 (2011).
15      Keinan, A., Mullikin, J. C., Patterson, N. & Reich, D. Accelerated genetic drift on chromosome X during the human dispersal out of Africa. *Nat. Genet.* **41**, 66–70 (2009).
16      Hvilsom, C. et al. Extensive X-linked adaptive evolution in central chimpanzees. *PNAS* **109**, 2054–2059 (2012).
17      Prado-Martinez, J. et al. Great ape genetic diversity and population history. *Nature* **499**, 471–475 (2013).
18      Davydov, E. V. et al. Identifying a high fraction of the human genome to be under selective constraint using GERP++. *PLoS Comput. Biol.* **6**, e1001025 (2010).
19      McDonald, J. H. & Kreitman, M. Adaptive protein evolution at the Adh locus in Drosophila. *Nature* **351**, 652–654 (1991).
20      Charlesworth, B., Coyne, J. A. & Barton, N. H. The relative rates of evolution of sex chromosomes and autosomes. *Am. Nat.* **130**, 113–146 (1987).
21      Rice, W. R. Sex chromosomes and the evolution of sexual dimorphism. *Evolution* **38**, 735–742 (1984).
22      Dean, R., Perry, J. C., Pizzari, T., Mank, J. E. & Wigby, S. Experimental evolution of a novel sexually antagonistic allele. *PLoS Genet.* **8**, e1002917 (2012).
23      Wang, P. J., McCarrey, J. R., Yang, F. & Page, D. C. An abundance of X-linked genes expressed in spermatogonia. *Nat. Genet.* **27**, 422–426 (2001).
24      Nielsen, R. et al. A Scan for positively selected genes in the genomes of humans and chimpanzees. *PLoS Biol.* **3**, e170 (2005).
25      Swanson, W. J., Nielsen, R. & Yang, Q. Pervasive adaptive evolution in mammalian fertilization proteins. *Mol. Biol. Evol.* **20**, 18–20 (2003).
26      Bersaglieri, T. et al. Genetic signatures of strong recent positive selection at the lactase gene. *Am. J. Hum. Genet.* **74**, 1111-1120 (2004).
27      Pan, J. et al. Inactivation of Nxf2 causes defects in male meiosis and age-dependent depletion of spermatogonia. *Dev. Biol.* **330**, 167–174 (2009).
28      Zhou, J. et al. Nxf3 is expressed in sertoli cells, but is dispensable for spermatogenesis. *Mol. Reprod. Dev.* **78**, 241–249 (2011).

Figure 1. **Diversity levels in and outside genes on autosomes and X chromosomes.** The phylogenetic relation among all investigated great apes is shown above. The nucleotide diversity of X chromosomes, autosomes, and the diversity ratio, (A) in exons, introns and intergenic regions, and (B) in windows binned according the distance from the nearest transcript (95% of confidence intervals from 1,000 bootstrapping iterations are shown).

Figure 2. **Diversity and proportion of singletons along X chromosomes.** For each species, the nucleotide diversity in non-overlapping 100 kb windows of called sequence is plotted in black color with corresponding values of the proportion of singletons among SNPs in the same windows in blue color. At the top of each panel red marks are placed for windows where the nucleotide diversity is less than 20% of the mean diversity for the X chromosome of the given species. This is not done for the eastern gorilla where there is too little data. The shaded areas with light green color are the ampliconic regions.

Figure 3. **Evidence for X chromosome specific sweeps**. A. Heatmap for shared regions of reduced diversity among species for the autosomes and the X chromosomes. The diagonal shows the percentage of the windows having less than 20% diversity of the mean and the off-diagonal squares show the percentage of windows satisfying this condition in both species compared. B. Boxplot comparing genetic differentiation in regions of reduced diversity to remaining regions in X chromosomes. Comparisons are among the chimpanzee subspecies and the two orang-utans. DeltaP is the average allele frequency difference for SNPs in the two type of regions compared.

Figure 4.**McDonald Kreitman tests for positive selection**. The inferred proportion of mutations fixed by positive selection, for different types of functional classes ('nonsyn' denotes nonsynonymous sites) for autosomes (red) and X chromosomes (blue) with significance, calculated by 100 times of bootstrapping. B. The relative ratio of the polymorphisms in the functionally constrained sequences to that in the neutrally evolving sequences. C. The proportion of singletons among SNPs (***, **, *, +, and ns denotes the p values with < 0.001, < 0.01, < 0.05, <0.10, and >= 0.10, respectively).

Table 1. **Expected diversity reduction by soft and hard sweeps.** Summary of expected length of reduced diversity due to soft sweeps and hard sweeps as a function of effective population size ($N$), selection coefficient ($s$), and the proportion of beneficial allele onset of selection ($p_0$). For details, see Supplementary Information.

| $N_e$ | $s$ | Hard sweep | | Soft sweep | | | |
| --- | --- | --- | --- | --- | --- | --- | --- |
| | | by 25% | by 75% | $p_0 = 0.01$ | | $p_0 = 0.1$ | |
| | | | | by 25% | by 75% | by 25% | by 75% |
| 10000 | 0.01 | 120kb | < 10kb | 120kb | < 10kb | 70kb | < 10kb |
| | 0.05 | 530kb | 100kb | 450kb | 80kb | 170kb | < 10kb |
| | 0.1 | 1Mb | 190kb | 730kb | 120kb | 190kb | < 10kb |
| | 0.2 | 1.8Mb | 370kb | 1Mb | 180kb | 220kb | < 10kb |
| | 0.5 | 5.4Mb | 1.2Mb | 1.7Mb | 250kb | 240kb | < 10kb |
| 50000 | 0.01 | 100kb | < 10kb | 80kb | < 10kb | 10kb | < 10kb |
| | 0.05 | 420kb | 90kb | 230kb | 40kb | 30kb | < 10kb |
| | 0.1 | 790kb | 160kb | 310kb | 40kb | 40kb | < 10kb |
| | 0.2 | 1.6Mb | 320kb | 400kb | 50kb | 40kb | < 10kb |
| | 0.5 | 4.3Mb | 870kb | 470kb | 60kb | 40kb | < 10kb |

Table 2. **Enriched ampliconic regions in the lower diversity on X chromosomes**
The observed number of amplicons overlapping windows of putative selective sweeps (O) by more than 30%, the expected number based on randomized location of amplicons with 10,000 replicates (E), the O/E ratio, and the P value for significant overlap.

|  | O | E | O/E | p-value |
|---|---|---|---|---|
| Bonobo | 5 | 0.5 | 10.0 | 0.0003 |
| Central Chimp | 8 | 0.6 | 13.3 | < 0.0001 |
| Eastern Chimp | 10 | 1.0 | 10.0 | < 0.0001 |
| Western Chimp | 11 | 1.7 | 6.5 | < 0.0001 |
| NC Chimp | 6 | 0.7 | 8.6 | < 0.0001 |
| Western Gorilla | 12 | 1.8 | 6.7 | < 0.0001 |
| Sumatran Orang | 9 | 1.4 | 6.4 | < 0.0001 |
| Bornean Orang | 6 | 2.5 | 2.4 | 0.0956 |

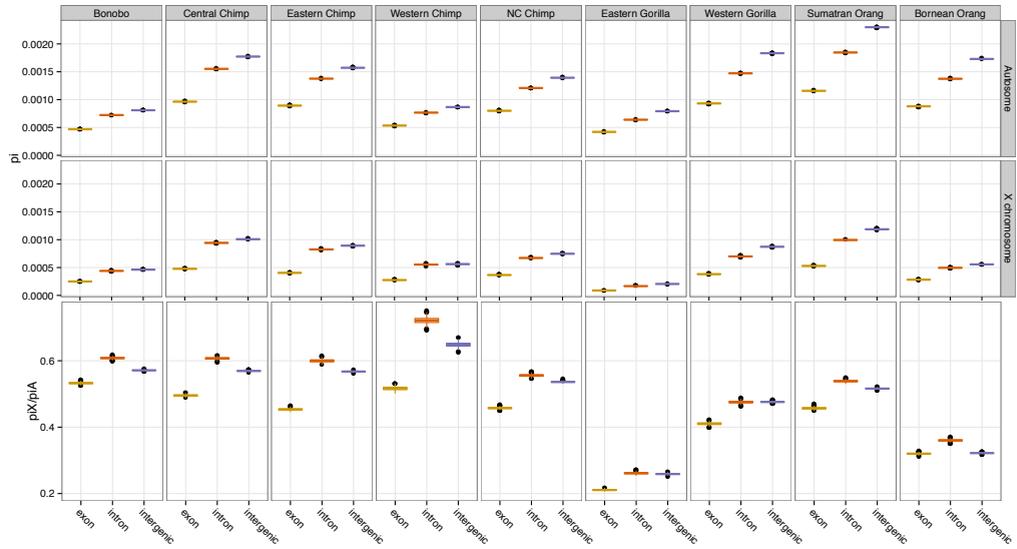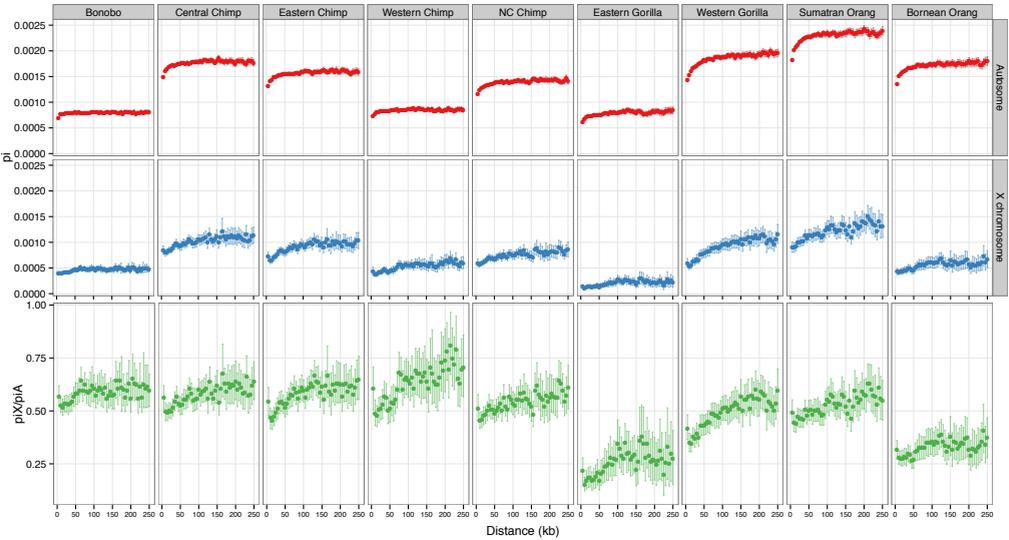

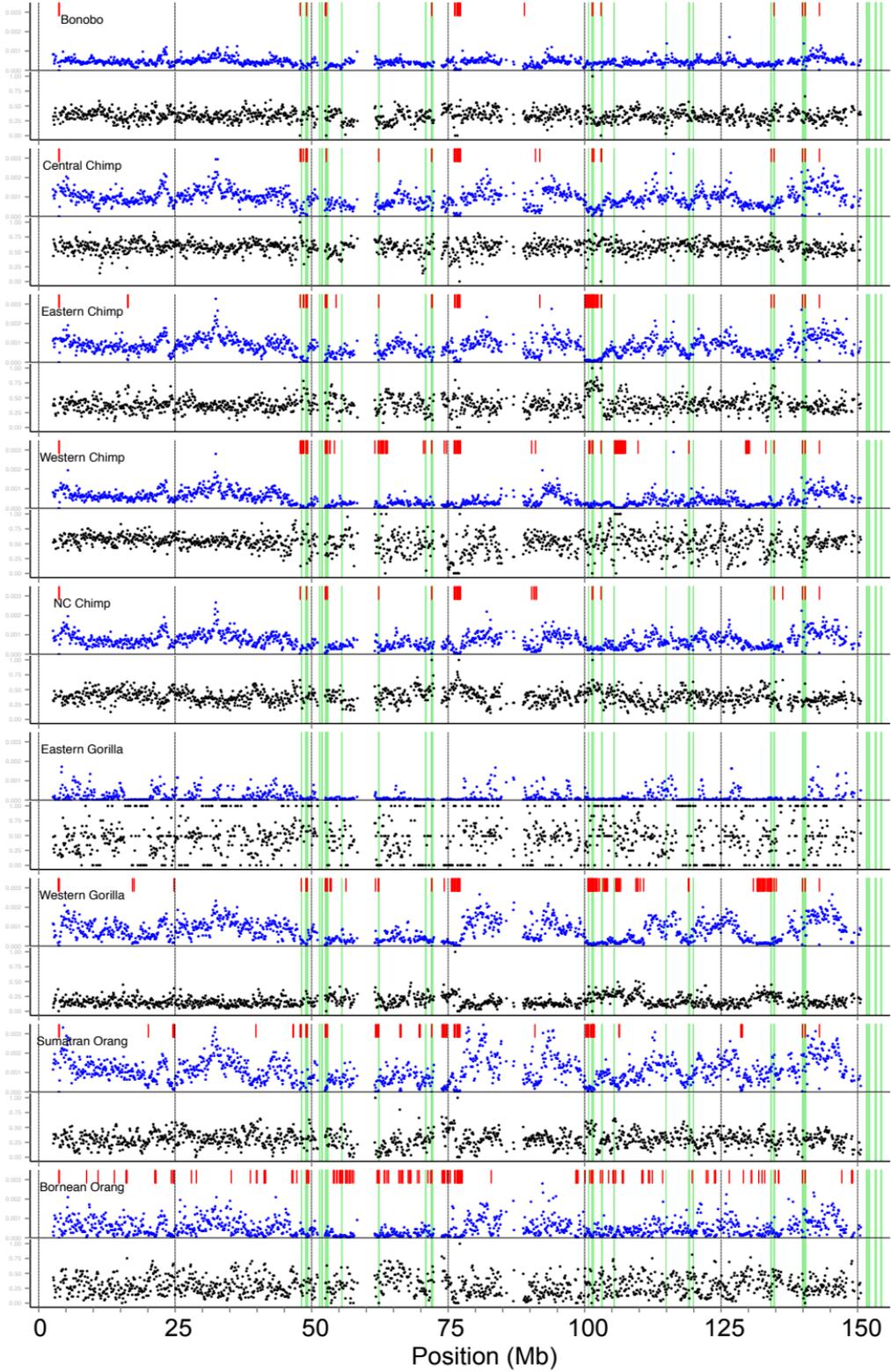

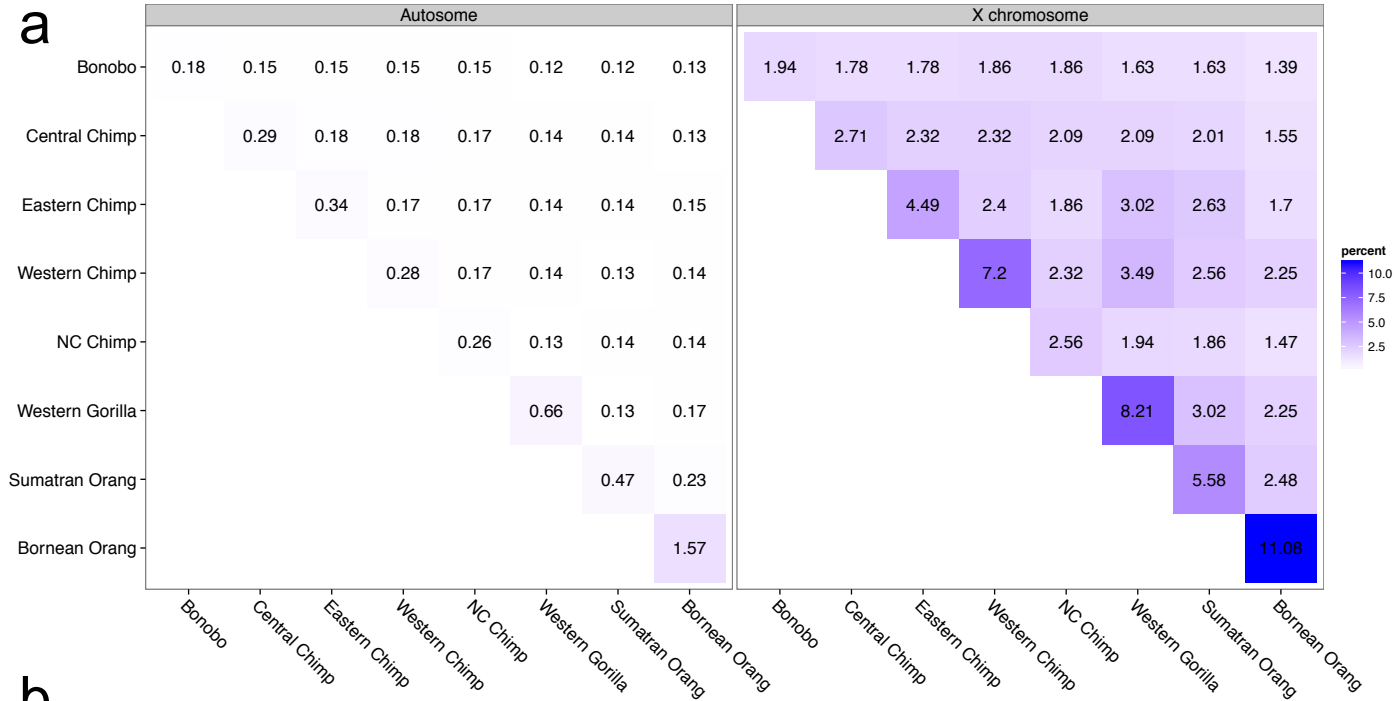

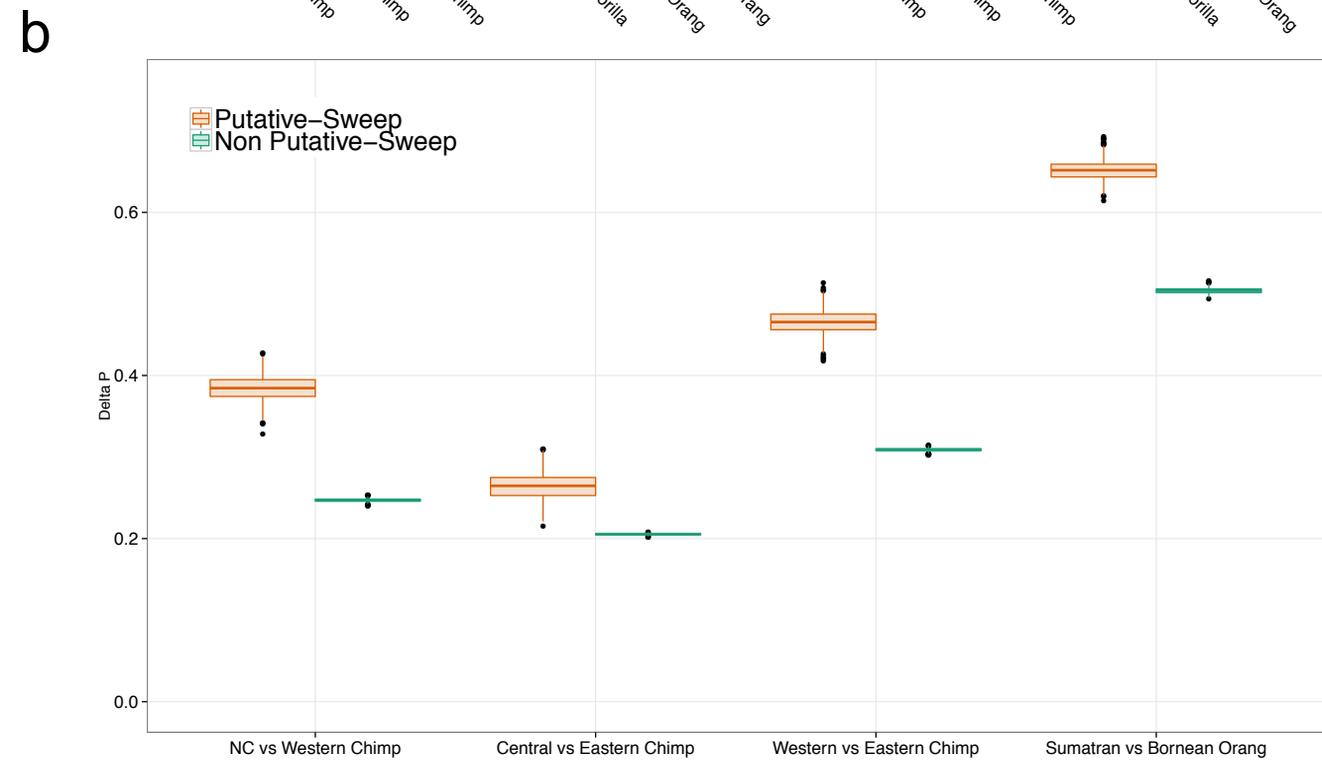

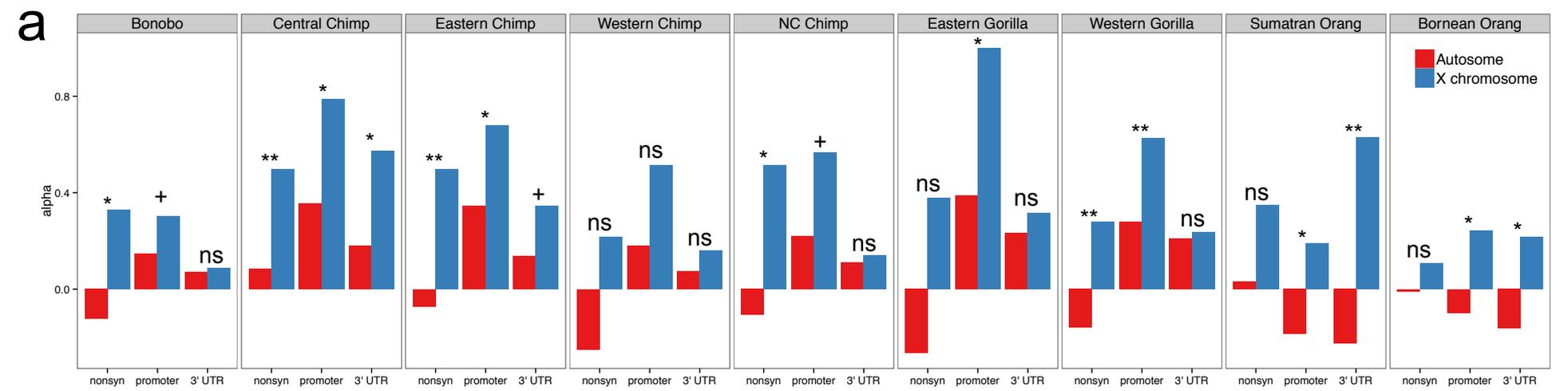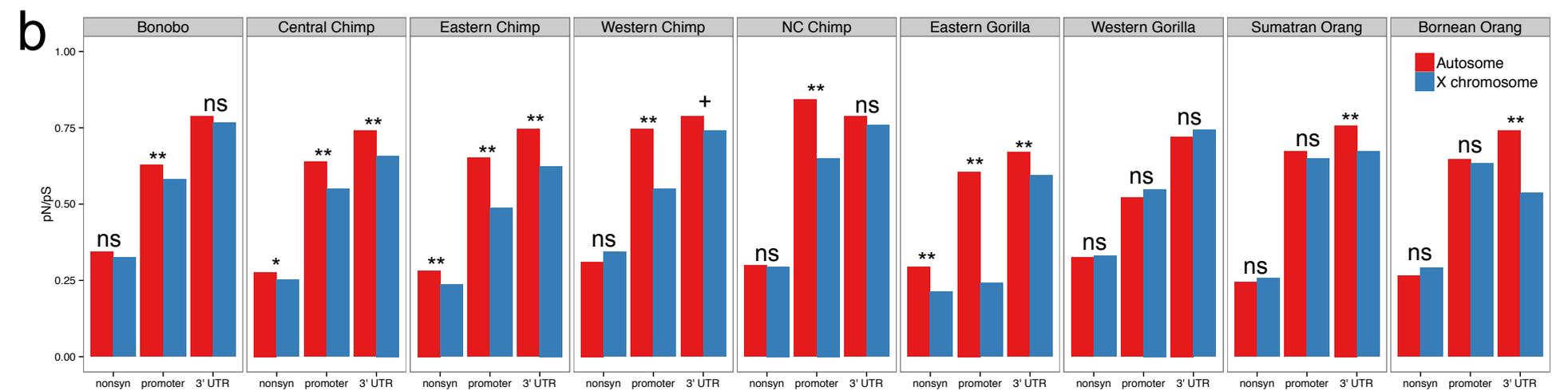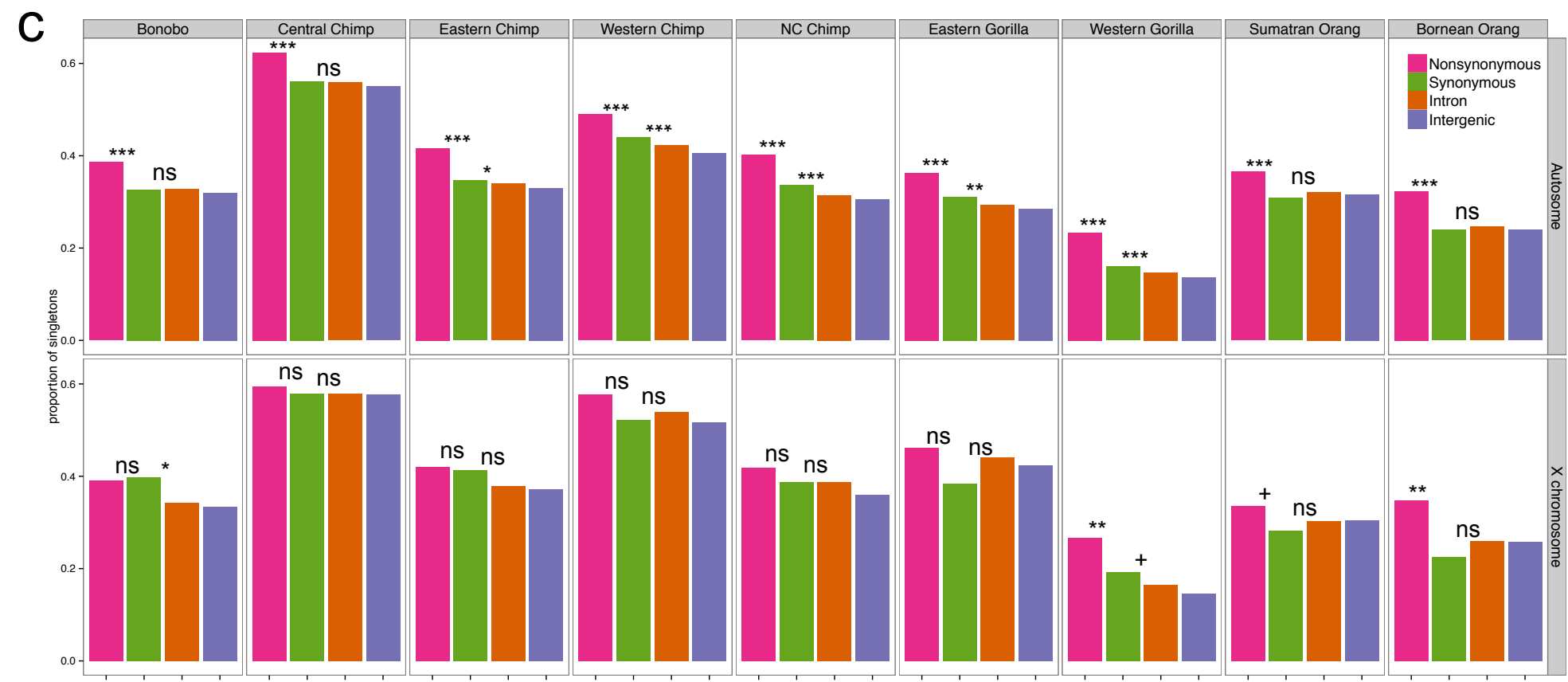

# Contents

# 1. Background selection

At a loci with no recombination the reduction in diversity is given by $\exp(-U/2s)$ where $U$ is the deleterious mutation rate and $s$ the selection intensity (Durrett 2002 Theorem 6.7; Charlesworth et al. 1993). Both $U$ and $s$ are unknown, but approximate $U$ as the product of the per-base mutation rate, $u$, the fraction of bases that are functional, $f$, the fraction of mutations of functional sites that will be deleterious, $d$, and the size of the locus under consideration, $L$:

$$U = u\,f\,d\,L.$$

Assuming that the mutation rate in the great apes is similar to humans, we have a mutation rate of $u$ = 1.2e-8 per generation (Kong et al. 2012). Further assuming that the fraction of functional sites is roughly proportional to the fraction of sequence that is exons (ENCODE results aside), $f$ is in the range 1-3%. To err on the side of too much reduction we will use $f = 5\%$. For $d$ we can, erring again on the side of too much reduction, use the expected fraction of non-synonymous to synonymous mutations $d = 70\%$. For $L$ it is harder to think of an obvious size for a recombination free region, considering that all regions recombine, but we can try to vary $L$ a few orders of magnitude.

For a range of $s$ values, the reduction is plotted below:

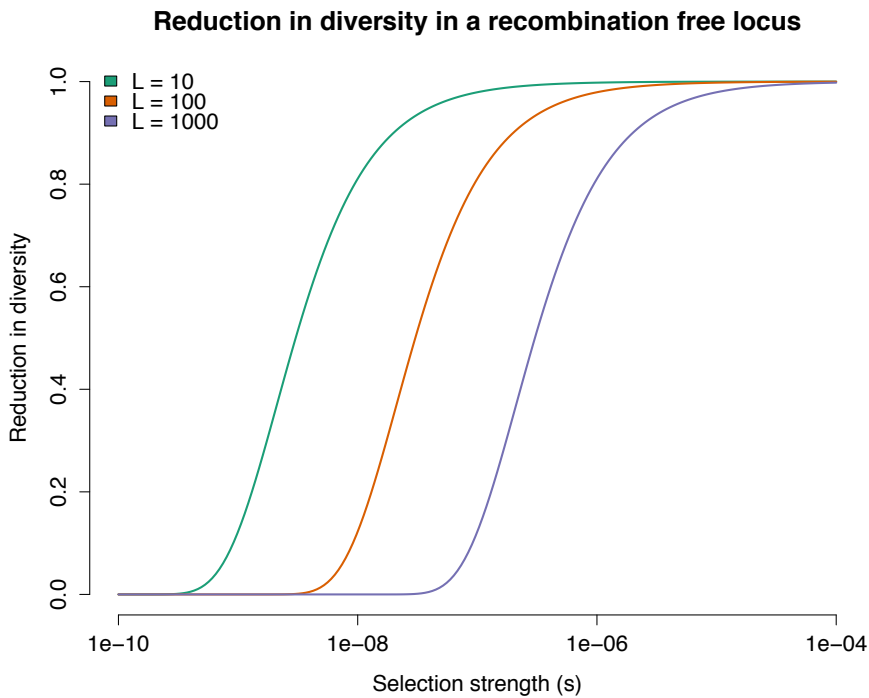

As apparent from the plot, it is possible to reduce the diversity significantly – to the point of practically no variation – provided that the selection is sufficiently weak. There is a limit to how weak we can set $s$, however, since $s < 1/Ne$ are effectively neutral (Charlesworth et al. 1993). In great apes $Ne$ is on the order of 1e4, and in ranges where $s$ is expected to have an effect the reduction is quite limited.

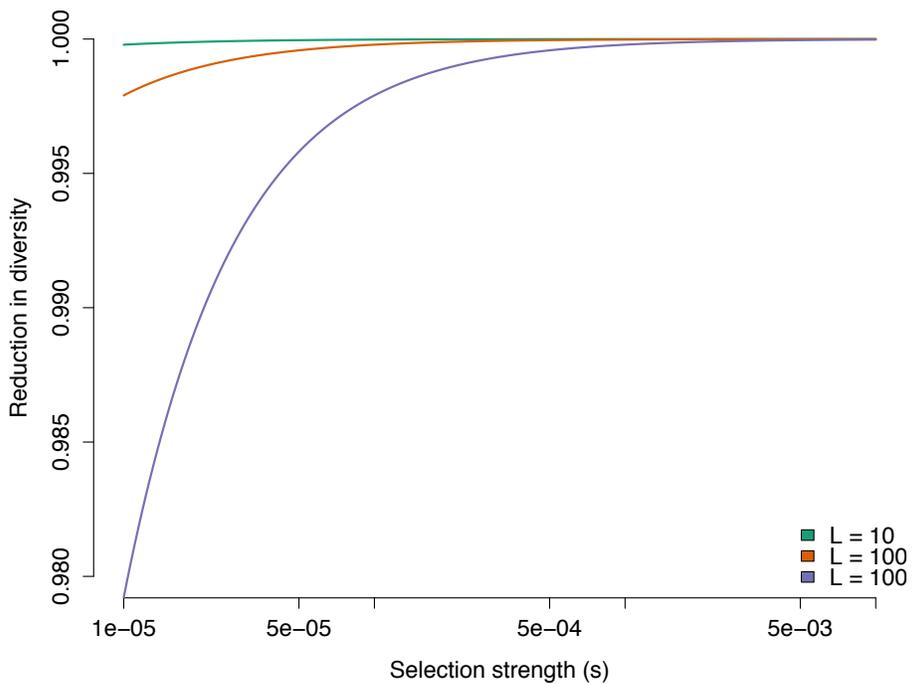

## Regions with recombination

While it is unrealistic to have very large regions without recombination, we can also include recombination in a model that assumes background selection in a region of size $L$ with both recombination and deleterious mutations. In such a region the reduction in diversity is given by $\exp(-U/(2s + R))$ where $U$ is given by $U = 2\,u\,f\,d\,L$ as before and where $R = 2\,r\,L$ is the total recombination in the region and $r$ is the per-basepair recombination rate (Durrett 2002 equation 6.24).

The mean recombination rate in humans is about 1 cM/Mb and presumably the rate is similar in the other great apes. The lower the recombination rate, the stronger the effect of background selection, and the regions we consider *do* have low recombination. However, they are reduced by less than an order of magnitude. Plotting the reduction for $r$ between 0.1 cM/Mb and 1.0 cM/Mb and varying region sizes, $L$, we get the plot below for $s$ = 1e-4 (around the lower limit for what we can reasonably assume to be under selection for the effective population sizes of the great apes):

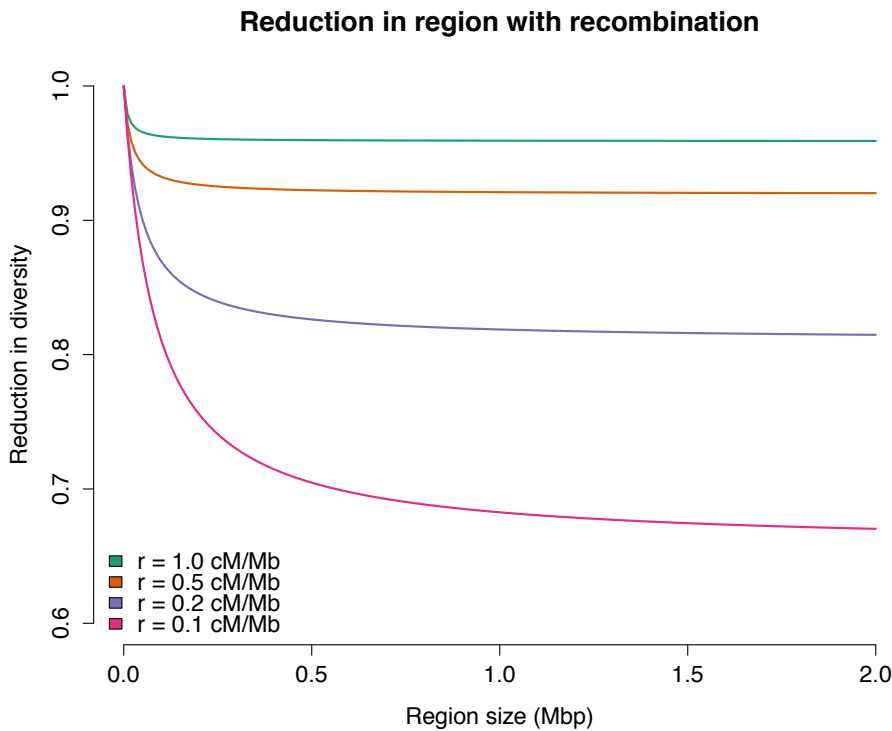

With smaller (but unrealistic) selection parameters diversity will drop faster as a function of $L$ and for larger $s$ it will drop slower. The plot below shows this for three different selection strengths and the four recombination rates shown above:

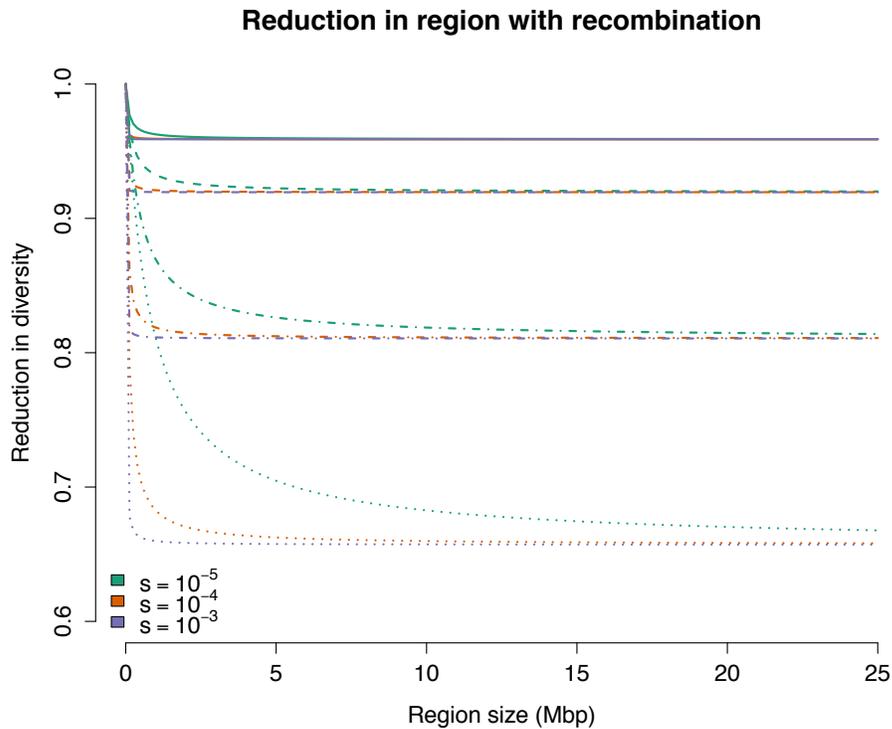

**Reduction in region with recombination**

While it is possible to reduce diversity considerably with background selection, it seems to require unrealistically small selection strengths or recombination rates. Requiring $s > 1/Ne$ for the effective population sizes seen in great apes makes it hard to reduce the diversity by more than a few percent unless the recombination rate is unusually low, and even for megabase large regions with a recombination rate below one tenth of the genomic average the diversity is still only reduced to about 70%. With a more realistic, although still low, recombination rate of 0.5 cM/Mb less than 10% of the diversity is expected to be lost.

**References**

Charlesworth, B., Morgan, M.T. & Charlesworth, D., 1993. The effect of deleterious mutations on neutral molecular variation. *Genetics*, 134(4), pp.1289–1303.

Durrett, R., 2002. *Probability Models for DNA Sequence Evolution*, Springer.

Kong, A. et al., 2012. Rate of de novo mutations and the importance of father's age to disease risk. *Nature*, 488(7412), pp.471–475.

# 2. Simulation of sweeps

To assess the potential effect of hard and soft sweeps on diversity on the X chromosome we performed a large number of simulations exploring combinations of selection coefficients $s$, effective population sizes $N$, and frequencies, $f$, of the selected variant at the onset of selection. We assume a discrete generation Wright-Fisher model and sample frequency trajectories of variants selected by coefficient $s$ using rejection sampling (rejecting trajectories where the selected variant does not go to fixation). Trajectories for hard sweeps thus begin with one and proceed to $2N * ¾$ by repeated binomial sampling with probability parameter $N_{mut}/(N_{mut} + (N - N_{mut})(1-s))$, where $N_{mut}$ is the number of selected variants in the previous generation. Trajectories for soft sweeps begin with an initial frequency $f$ of the selected variant and are prepended with a trajectory from 1 to $f * 2N * 3/4$ representing variant frequency prior to the onset of the selection.

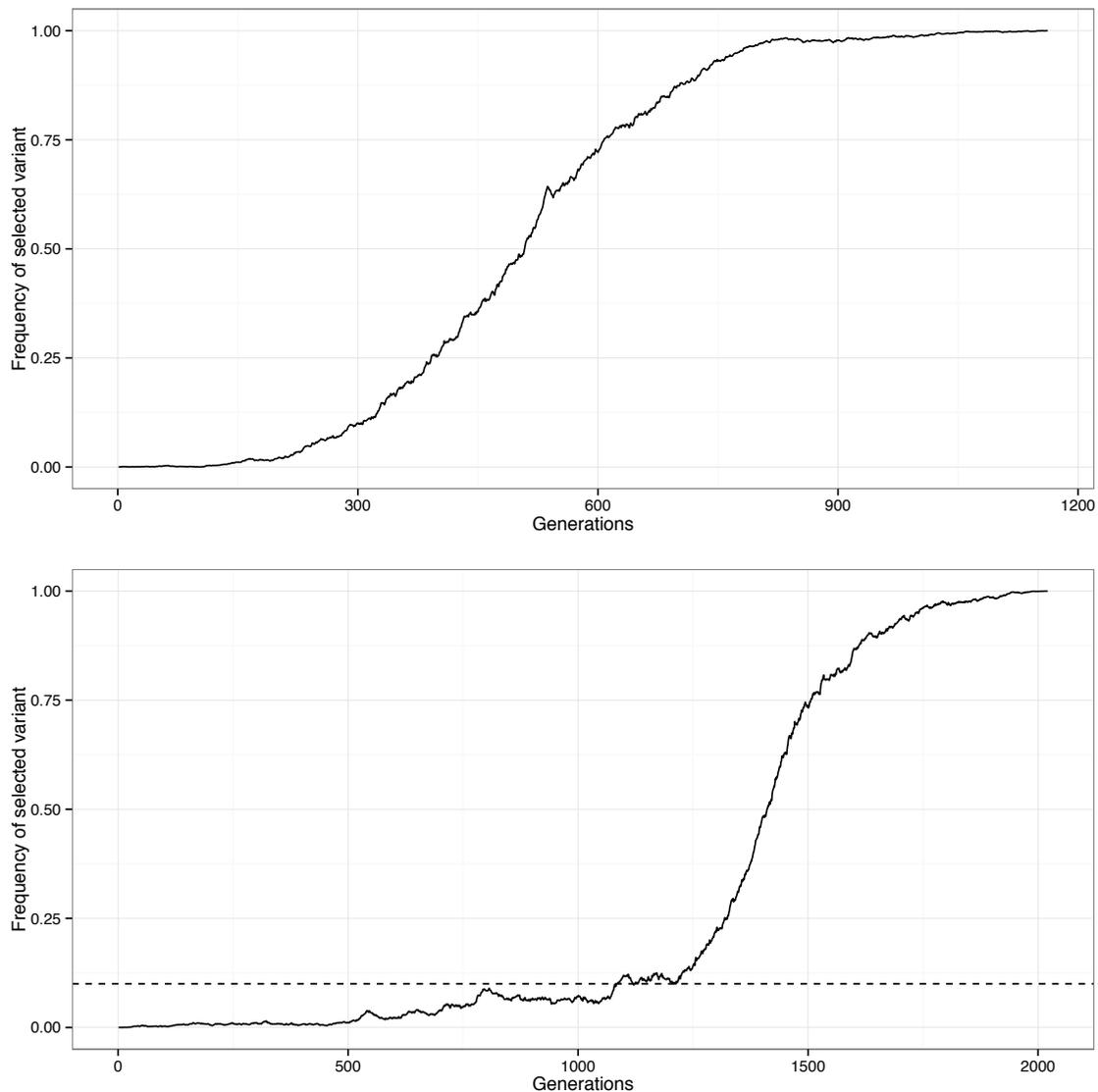

*Examples of trajectories with s=0.01 and N=10.000 for a hard sweep (top) and a soft sweep (bottom) with initial frequency 10%.*

We consider a sample of two sequences representing 10cM (10Mb assuming a recombination rate of 1 cM/Mb) with the selected variant at the 5' position (assuming that effect of a sweep is symmetric). To compute the time to the most recent common ancestor along the two sequences we simulate coalescence with recombination.

Given a recombination event the sequence to the right of the recombination point will become unlinked from the sweep. Lineages carrying the selected variant can only share an immediate ancestor with another lineage also carrying the variant. However, lineages not linked to the selected variant may coalesce with one that is and link the resulting lineage back into the sweep. The simulation proceeds until all sequence segments separated by recombination events have found a most recent commons ancestor (TMRCA). For each combination of parameters *s*, *N* and *f* we perform 1000 simulations and the mean TMRCA is computed in bins of 100kb. To more accurately measure the effect of week sweeps we performed 20,000 simulations with samples of two 0.5 cM sequences and computed the mean TMRCA is computed in bins of 10kb.

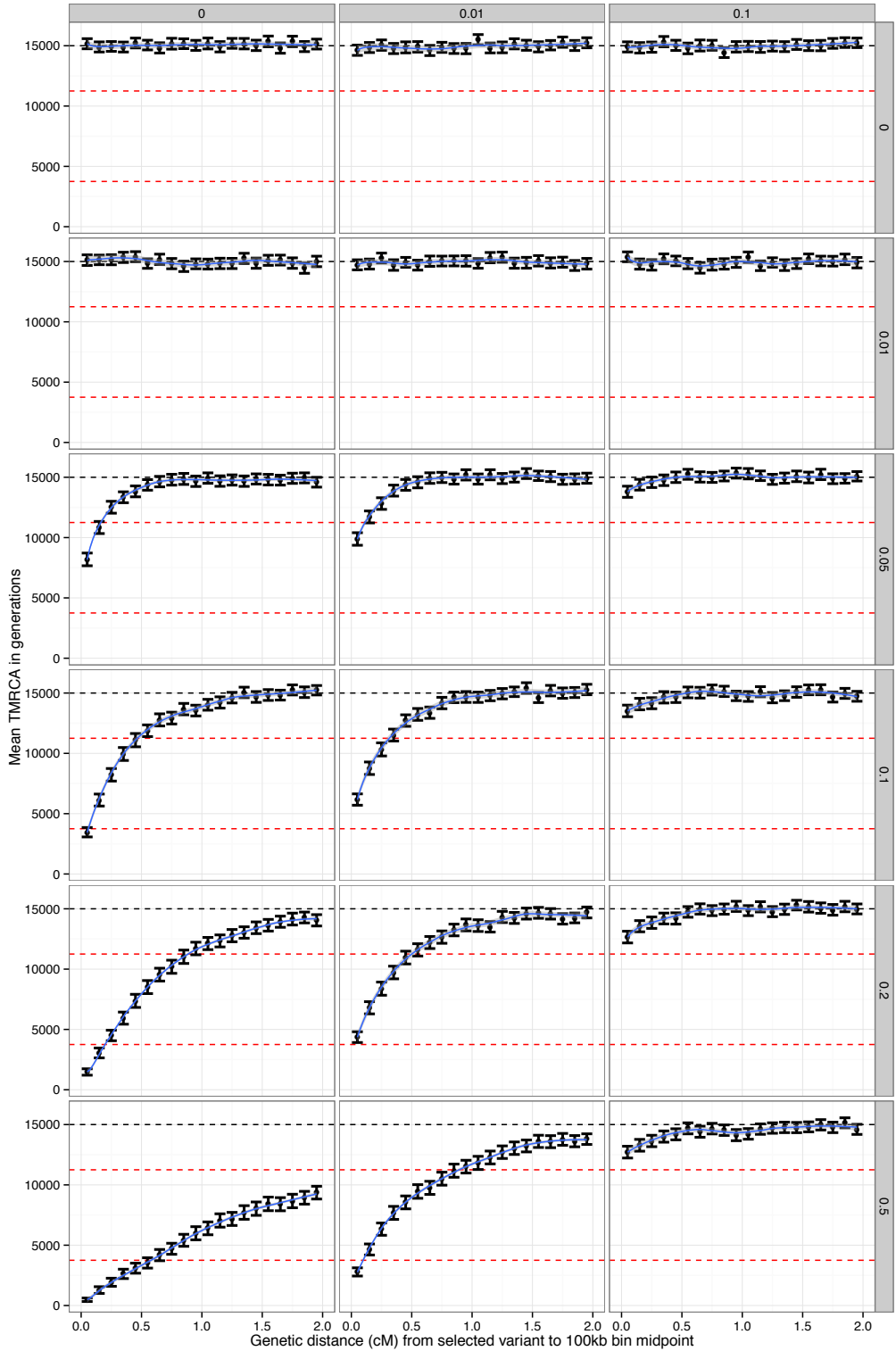

*Time to most recent common ancestor as function of genetic distance to selected variant for combinations of selection coefficients (right) and frequency of variant before onset of selection (top) A frequency of zero represents hard sweeps. The effective population size is set to 10.000. Each sub-plot is based on 1000 simulations. Horizontal black dashed lines show expected TMRCA without selection. Red dashed lines show 75% and 25% reductions in TMRCA.*

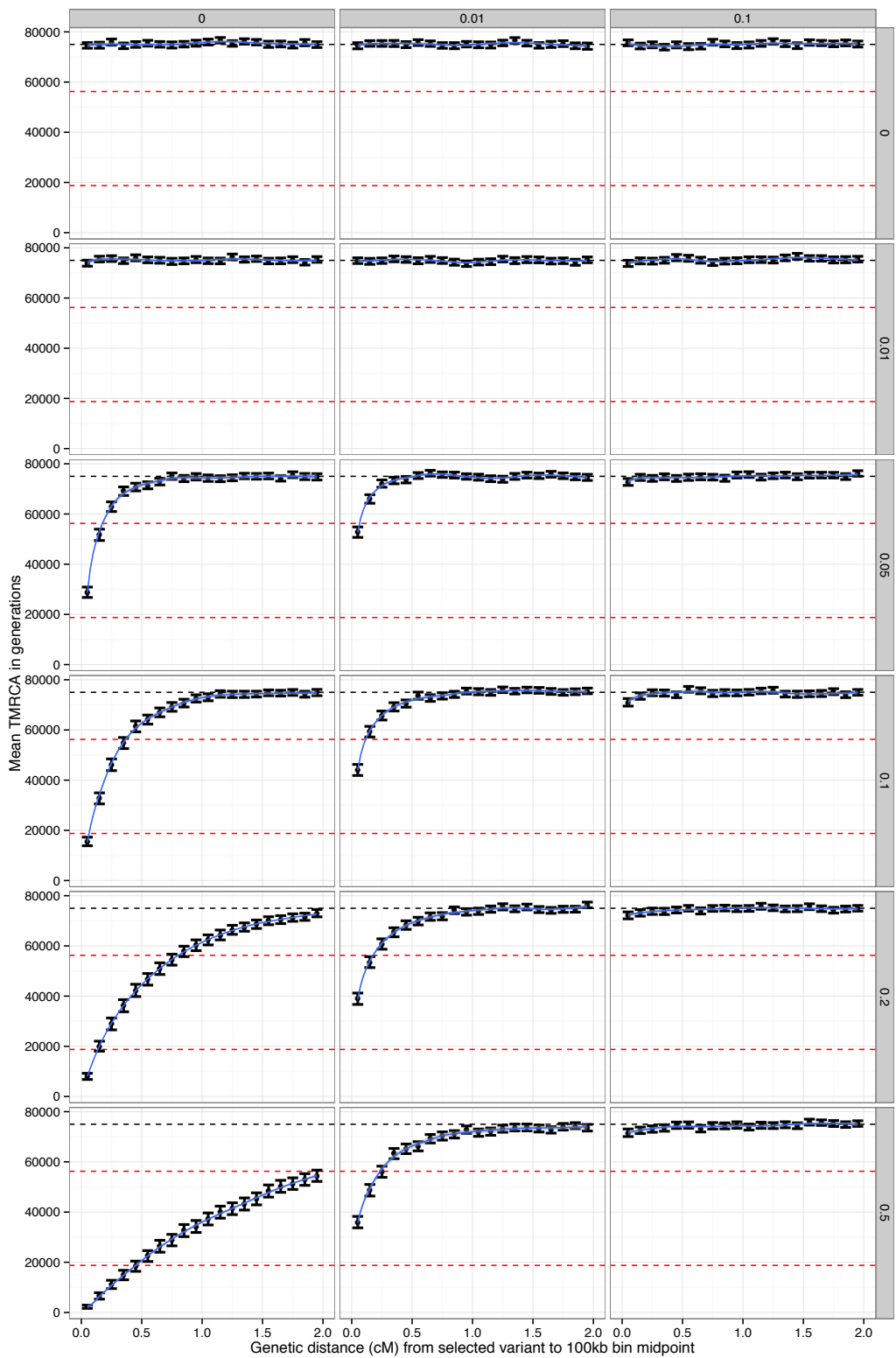

*Same as above but with an effective population size of 50.000*

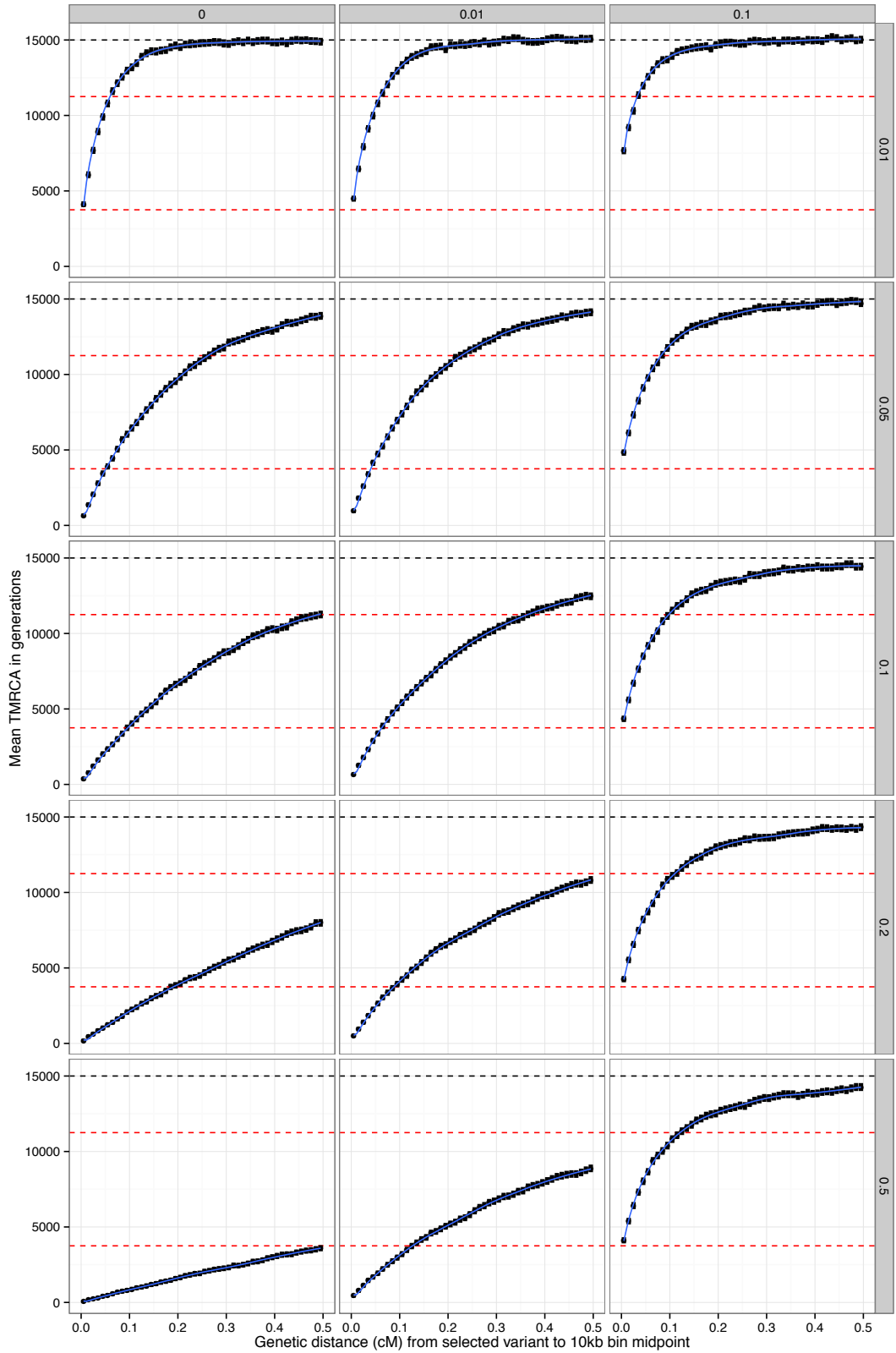

*Same as above but with N=10.000, 10kb bins and based on 20.000 simulations.*

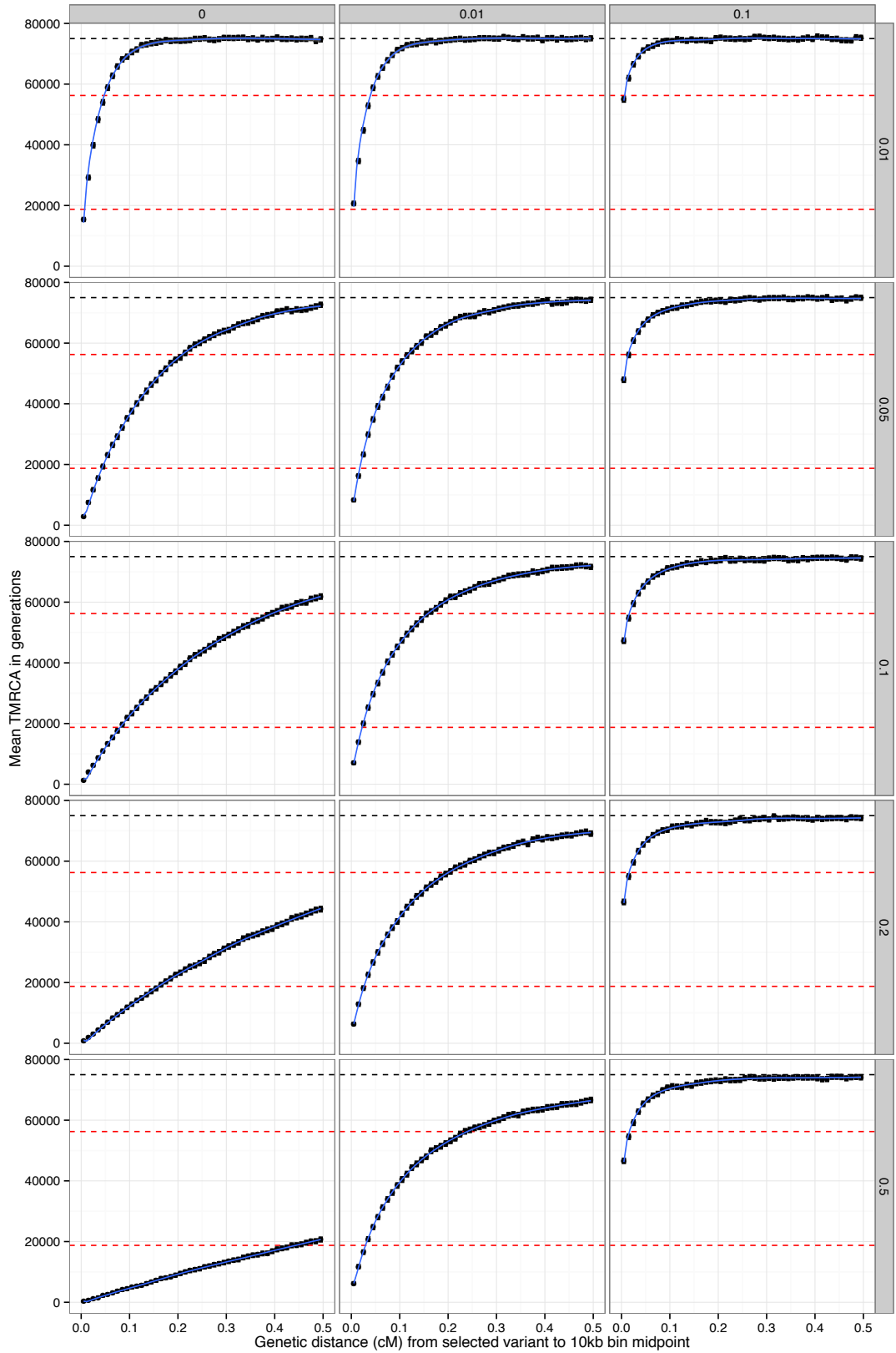

*Same as above but with N=50.000, 10kb bins and based on 20.000 simulations.*

# 3. Supplementary tables

Supplementary Table 1. **The information of taxa used in this study**. Human and Cross River Gorilla are not analyzed in this study because of lack of sequence quality and the small number of sequenced individuals

| Common name | Scientific name | Number of males | Number of females |
|---|---|---|---|
| Human | *Homo sapiens* | 10 | 0 |
| Bonobo | *Pan paniscus* | 2 | 11 |
| Central Chimpanzee | *Pan troglodytes troglodytes* | 1 | 3 |
| Eastern Chimpanzee | *Pan troglodytes schweinfurthii* | 2 | 4 |
| Western Chimpanzee | *Pan troglodytes verus* | 4 | 1 |
| NC Chimpanzee | *Pan troglodytes ellioti* | 4 | 6 |
| Eastern Lowland Gorilla | *Gorilla beringei graueri* | 2 | 1 |
| Cross River gorilla | *Gorilla gorilla diehli* | 0 | 1 |
| Western Lowland Gorilla | *Gorilla gorilla gorilla* | 4 | 23 |
| Sumatran Orang-utan | *Pongo abelii* | 1 | 4 |
| Bornean Oran-utan | *Pongo pygmaeus* | 1 | 4 |

Supplementary Table 2. **The information for each chromosome**. The length of chromosomes, the number of called positions, the number of 'N's in hg18, the number of analyzed positions, the length of exons, the length of introns, the length of intergenic sequences, and the length of repetitive sequences.

| chromosome | Length | Called | N | Analyzed | exon | intron | intergenic | repeat |
|---|---|---|---|---|---|---|---|---|
| chr1 | 247,249,719 | 140,388,359 | 11,022,139 | 129,366,220 | 3725987 | 57,710,683 | 67,929,550 | 63,331,551 |
| chr2 | 242,951,149 | 159,786,334 | 1,815,867 | 157,970,467 | 3102934 | 64,182,178 | 90,685,355 | 73,077,586 |
| chr3 | 199,501,827 | 137,263,385 | 3,236,108 | 134,027,277 | 2663488 | 59,274,923 | 72,088,866 | 64,147,653 |
| chr4 | 191,273,063 | 125,307,894 | 2,413,565 | 122,894,329 | 1791866 | 42,546,230 | 78,556,233 | 60,092,058 |
| chr5 | 180,857,866 | 117,798,768 | 1,753,490 | 116,045,278 | 1929930 | 40,678,652 | 73,436,696 | 55,224,045 |
| chr6 | 170,899,992 | 112,989,085 | 1,934,088 | 111,054,997 | 2261789 | 44,222,139 | 64,571,069 | 52,197,249 |
| chr7 | 158,821,424 | 90,489,207 | 905,663 | 89,583,544 | 1729706 | 40,306,647 | 47,547,191 | 42,114,343 |
| chr8 | 146,274,826 | 97,875,966 | 1,705,835 | 96,170,131 | 1465776 | 37,445,801 | 57,258,554 | 46,216,906 |
| chr9 | 140,273,252 | 70,546,810 | 11,377,768 | 59,169,042 | 1067407 | 23,296,189 | 34,805,446 | 28,619,424 |
| chr10 | 135,374,737 | 82,675,615 | 1,752,358 | 80,923,257 | 1691628 | 37,303,495 | 41,928,134 | 37,689,528 |
| chr11 | 134,452,384 | 82,810,198 | 912,079 | 81,898,119 | 2295723 | 34,886,606 | 44,715,790 | 39,802,827 |
| chr12 | 132,349,534 | 89,794,103 | 746,894 | 89,047,209 | 2235180 | 38,555,962 | 48,256,067 | 43,914,766 |
| chr13 | 114,142,980 | 65,460,180 | 12,706,750 | 52,753,430 | 674870 | 17,771,612 | 34,306,948 | 24,713,812 |
| chr14 | 106,368,585 | 59,867,289 | 12,084,954 | 47,782,335 | 988962 | 18,881,050 | 27,912,323 | 23,310,290 |
| chr15 | 100,338,915 | 48,018,840 | 10,416,703 | 37,602,137 | 1099367 | 19,902,799 | 16,599,971 | 18,105,842 |
| chr16 | 88,827,254 | 45,614,288 | 7,339,821 | 38,274,467 | 1131308 | 16,438,553 | 20,704,606 | 18,914,047 |
| chr17 | 78,774,742 | 44,740,265 | 344,727 | 44,395,538 | 1928902 | 22,957,555 | 19,509,081 | 21,116,562 |
| chr18 | 76,117,153 | 54,829,760 | 1,020,158 | 53,809,602 | 842000 | 21,073,418 | 31,894,184 | 24,237,707 |
| chr19 | 63,811,651 | 25,252,570 | 5,482,029 | 19,770,541 | 1343331 | 9,510,869 | 8,916,341 | 11,209,634 |
| chr20 | 62,435,964 | 42,181,268 | 1,697,534 | 40,483,734 | 1002386 | 17,759,416 | 21,721,932 | 20,095,584 |
| chr21 | 46,944,323 | 22,262,601 | 8,484,723 | 13,777,878 | 146884 | 4,233,354 | 9,397,640 | 6,461,180 |
| chr22 | 49,691,432 | 17,683,320 | 7,850,123 | 9,833,197 | 397094 | 5,522,298 | 3,913,805 | 4,855,551 |
| chrX | 154,913,754 | 84,045,787 | 1,447,736 | 82,598,051 | 1381543 | 25,172,939 | 56,292,588 | 47,832,071 |
| Sum | 3,022,646,526 | 1,817,681,892 | 108,451,112 | 1,709,230,780 | 36898061 | 699,633,368 | 972,948,370 | 827,280,216 |

Supplementary Table 3. **The number of SNPs in each chromosome**. Abbreviations: B: Bonobo, CC: Central Chimpanzee, EC: Eastern Chimpanzee, WC: Western Chimpanzee, NC: Nigeria-Cameron Chimpanzee, ELG: Eastern Lowland Gorilla, WLG: Western Lowland Gorilla, SO: Sumatran Orang-utan, BO: Bornean Orang-utan

| Chromosome | B | CC | EC | WC | NC | ELG | WLG | SO | BO |
|---|---|---|---|---|---|---|---|---|---|
| chr1 | 424,498 | 567,242 | 542,503 | 322,889 | 584,589 | 200,228 | 839,961 | 675,572 | 491,463 |
| chr2 | 546,190 | 725,938 | 691,867 | 364,623 | 773,765 | 227,511 | 1,078,913 | 882,050 | 626,145 |
| chr3 | 449,245 | 603,137 | 580,123 | 306,598 | 661,603 | 205,894 | 901,393 | 731,672 | 521,566 |
| chr4 | 436,234 | 589,920 | 580,985 | 258,196 | 635,542 | 201,786 | 900,385 | 767,630 | 545,484 |
| chr5 | 401,196 | 531,972 | 506,728 | 250,910 | 571,390 | 168,936 | 812,657 | 676,590 | 484,245 |
| chr6 | 389,282 | 506,684 | 478,105 | 299,140 | 522,570 | 169,625 | 763,649 | 628,528 | 454,737 |
| chr7 | 316,236 | 421,628 | 401,673 | 241,802 | 439,976 | 137,988 | 633,505 | 490,258 | 358,030 |
| chr8 | 339,595 | 463,801 | 448,474 | 229,124 | 495,923 | 159,246 | 699,566 | 595,415 | 423,535 |
| chr9 | 218,091 | 296,780 | 277,963 | 155,102 | 313,504 | 88,202 | 409,672 | 356,264 | 237,744 |
| chr10 | 291,067 | 377,523 | 361,451 | 168,253 | 408,051 | 121,426 | 586,855 | 500,396 | 345,845 |
| chr11 | 277,289 | 364,488 | 349,119 | 193,012 | 387,594 | 104,227 | 562,917 | 440,008 | 313,927 |
| chr12 | 304,515 | 399,642 | 380,373 | 235,340 | 419,126 | 132,323 | 595,233 | 475,470 | 315,983 |
| chr13 | 173,696 | 249,911 | 234,493 | 141,599 | 261,480 | 85,820 | 375,697 | 328,049 | 226,344 |
| chr14 | 164,119 | 215,313 | 196,722 | 136,510 | 223,173 | 65,812 | 324,981 | 269,067 | 193,813 |
| chr15 | 122,701 | 159,303 | 151,526 | 84,585 | 168,148 | 49,309 | 234,959 | 194,502 | 143,105 |
| chr16 | 140,892 | 183,537 | 174,339 | 111,065 | 193,309 | 62,938 | 262,367 | 220,235 | 161,004 |
| chr17 | 152,658 | 197,759 | 191,374 | 85,648 | 215,348 | 62,693 | 289,729 | 251,154 | 177,623 |
| chr18 | 194,627 | 255,293 | 242,774 | 110,867 | 275,551 | 90,670 | 398,617 | 329,784 | 231,799 |
| chr19 | 75,606 | 91,036 | 87,777 | 59,351 | 96,744 | 31,675 | 141,406 | 115,153 | 85,403 |
| chr20 | 146,723 | 188,109 | 181,554 | 101,816 | 201,932 | 49,729 | 287,725 | 231,778 | 159,121 |
| chr21 | 54,635 | 75,320 | 69,897 | 49,807 | 77,445 | 21,348 | 124,436 | 110,945 | 73,153 |
| chr22 | 35,902 | 44,152 | 43,098 | 22,443 | 46,700 | 16,017 | 65,227 | 58,826 | 39,925 |
| chrX | 165,313 | 207,621 | 199,944 | 101,441 | 214,343 | 25,262 | 271,450 | 239,503 | 111,351 |
| Sum | 5,820,310 | 7,716,109 | 7,342,862 | 4,030,121 | 8,187,806 | 2,478,665 | 11,561,300 | 9,568,849 | 6,721,345 |

Supplementary Table 4. **The higher ratio of nonsynonymous to synonymous substitutuion in the reduced diversity.** The number of substitution from the ancestral great apes for African great apes and from hg18 for orang-utans was compared between the regions of the reduced diversity from multiple major lineages of the great apes and the rest of X chromosomes. The p values were calculated by the one-tailed fisher's exact test.

| species | reduced diversity | | the rest | | odd ratio | p value |
|---|---|---|---|---|---|---|
| | Nonsynonymous | synonymous | nonsynonymous | synonymous | | |
| Bonobo | 54 | 35 | 603 | 585 | 1.4963 | 0.0446 |
| Central Chimp | 56 | 37 | 580 | 545 | 1.4218 | 0.0666 |
| Eastern Chimp | 56 | 40 | 587 | 560 | 1.3353 | 0.1070 |
| Western Chimp | 59 | 40 | 608 | 573 | 1.3897 | 0.0735 |
| NC Chimp | 55 | 37 | 588 | 549 | 1.3875 | 0.0832 |
| Eastern Gorilla | 61 | 68 | 724 | 855 | 1.0593 | 0.4112 |
| Western Gorilla | 59 | 65 | 652 | 777 | 1.0817 | 0.3719 |
| Sumatran Orang | 58 | 58 | 668 | 887 | 1.3276 | 0.0843 |
| Bornean Orang | 65 | 66 | 705 | 935 | 1.3059 | 0.0839 |

Supplementary Table 5. **Test for enrichment of reproduction in the low diversity regions.** The one-tailed fisher's exact test was performed to test the enrichment of testis-specific genes and ovary-specific genes in the regions with low diversity compared with the rest X chromosomes. 'Shared' denotes the regions in which the low diversity was observed from multiple major lineage among great apes.

| Category | species | reproduction/ampliconic genes within putative sweeps | reproduction/ampliconic genes outside putative sweeps | non-reproduction/ampliconic genes within putative sweeps | non-reproduction/ampliconic genes outside putative sweeps | Odds ratio | P values |
|---|---|---|---|---|---|---|---|
| Testis | Bonobo | 3 | 95 | 4 | 573 | 4.51 | 0.067 |
| | NC chimp | 2 | 96 | 6 | 571 | 1.98 | 0.328 |
| | Eastern chimp | 5 | 93 | 35 | 542 | 0.83 | 0.716 |
| | Central chimp | 5 | 93 | 10 | 567 | 3.04 | 0.053 |
| | Western chimp | 7 | 91 | 29 | 548 | 1.45 | 0.258 |
| | Eastern gorilla | 21 | 77 | 106 | 471 | 1.21 | 0.278 |
| | Wesern gorilla | 8 | 90 | 33 | 544 | 1.46 | 0.232 |
| | Sumatran Orang | 7 | 91 | 16 | 561 | 2.69 | 0.037 |
| | Bornean Orang | 3 | 95 | 37 | 540 | 0.46 | 0.948 |
| | Shared | 6 | 92 | 24 | 553 | 1.50 | 0.260 |
| Ovary | Bonobo | 0 | 20 | 7 | 649 | 0.00 | 1.000 |
| | NC chimp | 0 | 20 | 8 | 648 | 0.00 | 1.000 |
| | Eastern chimp | 4 | 16 | 36 | 620 | 4.29 | 0.026 |
| | Central chimp | 0 | 20 | 15 | 641 | 0.00 | 1.000 |
| | Western chimp | 1 | 19 | 35 | 621 | 0.93 | 0.671 |
| | Eastern gorilla | 5 | 15 | 122 | 534 | 1.46 | 0.317 |
| | Wesern gorilla | 5 | 15 | 36 | 620 | 5.71 | 0.005 |
| | Sumatran Orang | 0 | 20 | 23 | 633 | 0.00 | 1.000 |
| | Bornean Orang | 2 | 18 | 38 | 618 | 1.80 | 0.334 |
| | Shared | 3 | 17 | 27 | 629 | 4.10 | 0.054 |

Supplementary Table 6. **Test for enrichment of ampliconic genes in the low diversity regions.** The one-tailed fisher's exact test was performed to test the enrichment of ampliconic genes in the regions with low diversity compared with the rest X chromosomes. 'Shared' denotes the regions in which the low diversity was observed from multiple major lineage among great apes.

| species | ampliconic genes within putative sweeps | ampliconic genes outside putative sweeps | amplliconic genes within putative sweeps | ampliconic genes outside putative sweeps | Odds ratio | p.value |
|---|---|---|---|---|---|---|
| Bonobo | 4 | 40 | 14 | 657 | 4.67 | 0.020 |
| NC chimp | 4 | 40 | 15 | 656 | 4.36 | 0.025 |
| Eastern chimp | 4 | 40 | 53 | 618 | 1.17 | 0.473 |
| Central chimp | 5 | 39 | 26 | 645 | 3.17 | 0.036 |
| Western chimp | 7 | 37 | 46 | 625 | 2.57 | 0.037 |
| Wesern gorilla | 7 | 37 | 51 | 620 | 2.30 | 0.056 |
| Sumatran Orang | 8 | 36 | 32 | 639 | 4.42 | 0.002 |
| Bornean Orang | 1 | 43 | 51 | 620 | 0.28 | 0.968 |
| Shared | 7 | 37 | 40 | 631 | 2.98 | 0.020 |

# 4. Supplementary figures

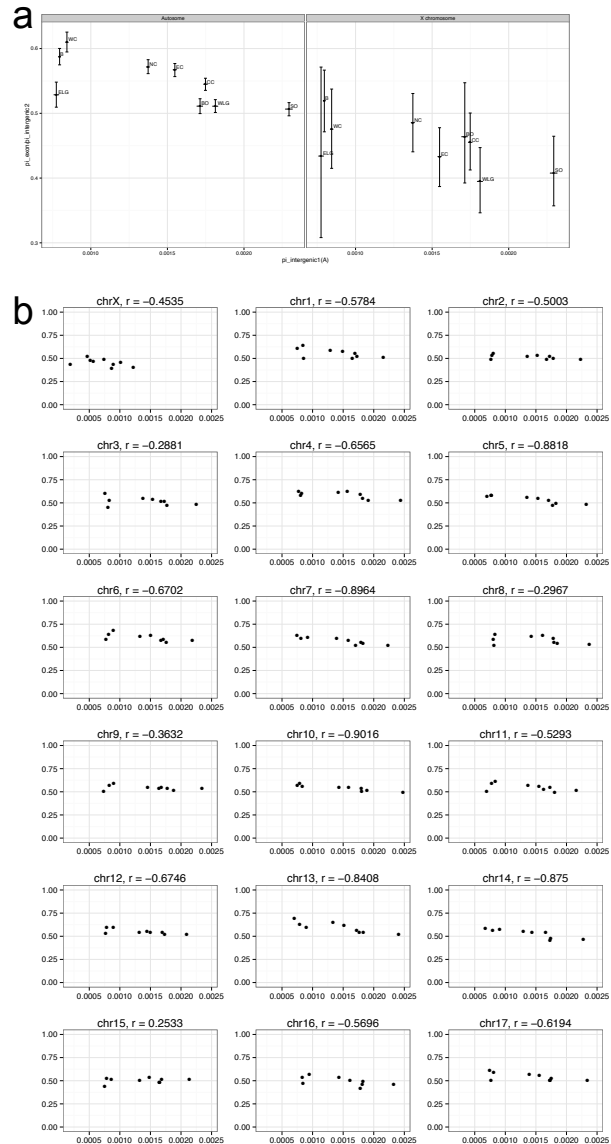

Supplementary Figure 1. **The relationship between reduction of diversity in exons and the intergenic diversity**. A. Both autosomes and X chromosomes have negative correlations between the relative reduction of exonic diversity to intergenic diversity and the autosomal diversity ($r = -0,69$, $p = 0.0410$ for autosomes; $r = -0.63$, $p = 0.0707$ for X chromosomes). B. For each chromosome, correlation between the relative reductions of exonic diversity with intergenic diversity of corresponding chromosome was also calculated. X chromosome has comparable correlation coefficient with autosomes. Please note that the intergenic sequences are divided into two groups to avoid autocorrelation and these two are used to estimate the relative reduction of the exonic diversity (y axis) and to infer the population size (x axis), respectively.

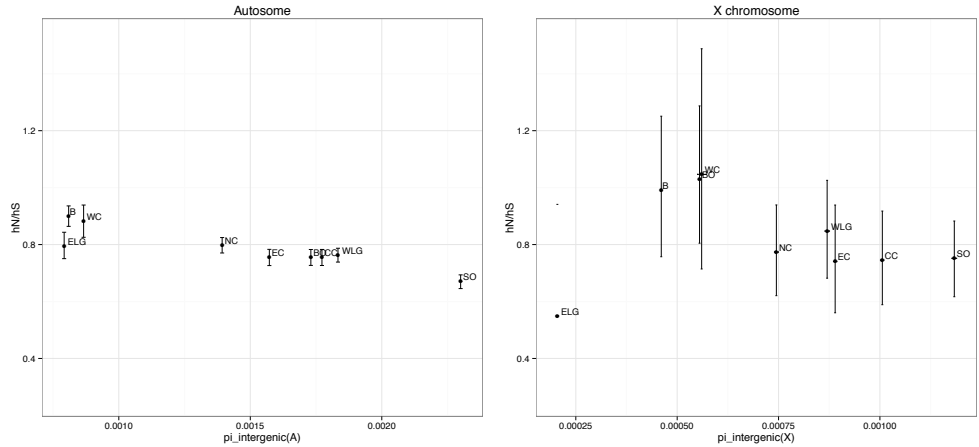

Supplementary Figure 2. **The relationship between the proportion of non-synonymous polymorphisms to synonymous polymorphisms and the intergenic diversity**. The ratio of nonsynonymous to synonymous heterozygosity is negatively correlated with the intergenic pi values in autosomes ($r = -0.89$, $p = 0.0011$), implying more efficient purifying selection with a lager population size. For X chromosomes, the correlation is not significant ($r = -0.10$, $p = 0.8046$). But when Eastern Lowland Gorilla is excluded, the correlation is strongly significant ($r = -0.87$, $p = 0.0065$).

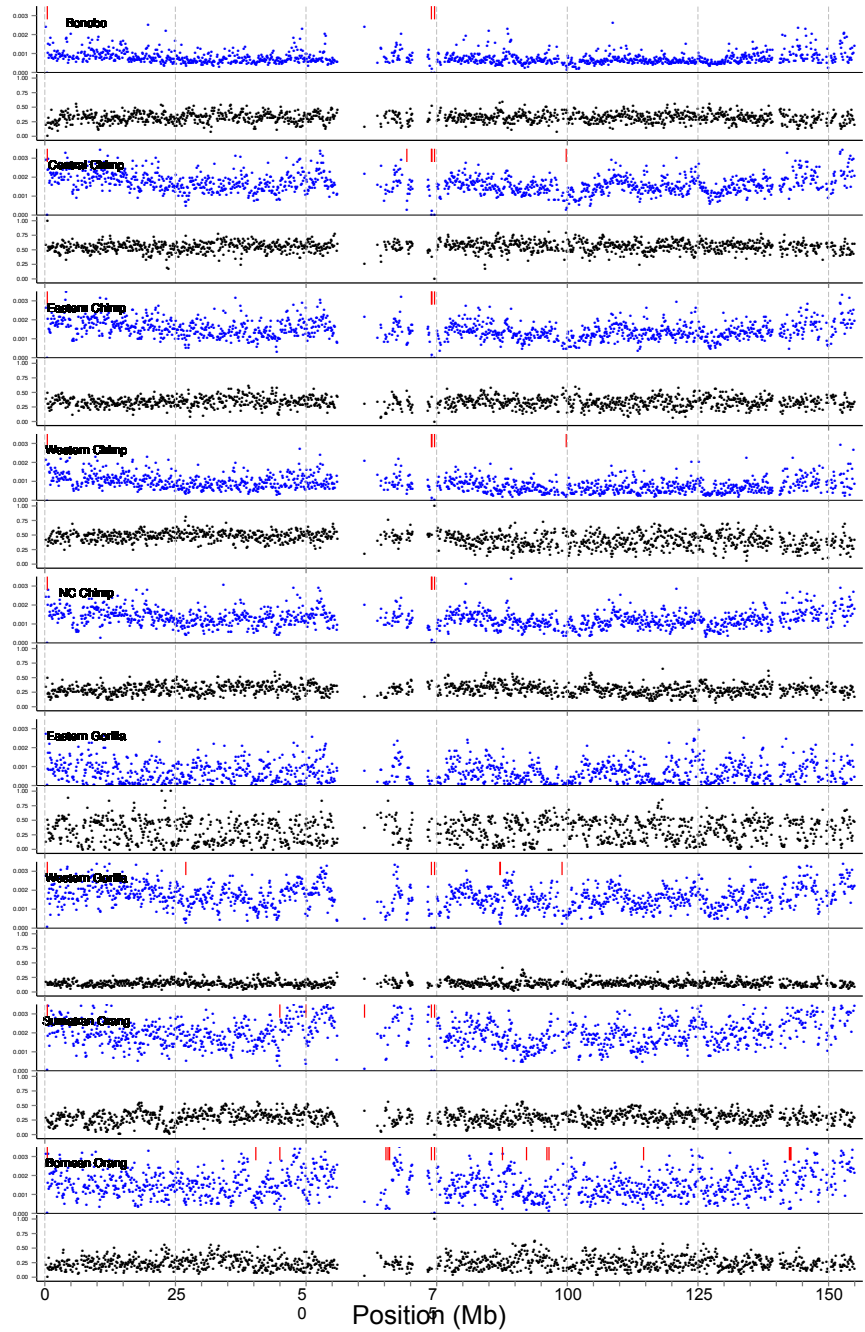

Supplementary Figure 3a. **The pattern of diversity and site-frequency spectrum in an autosome.** The pi value (blue) and the proportion of singletons (black) along chromosome 7, which has a comparable size with X chromosomes. Putative candidate sweeps are identified from the lower pi values than 20% of mean of the chromosomes and denoted as red bars.

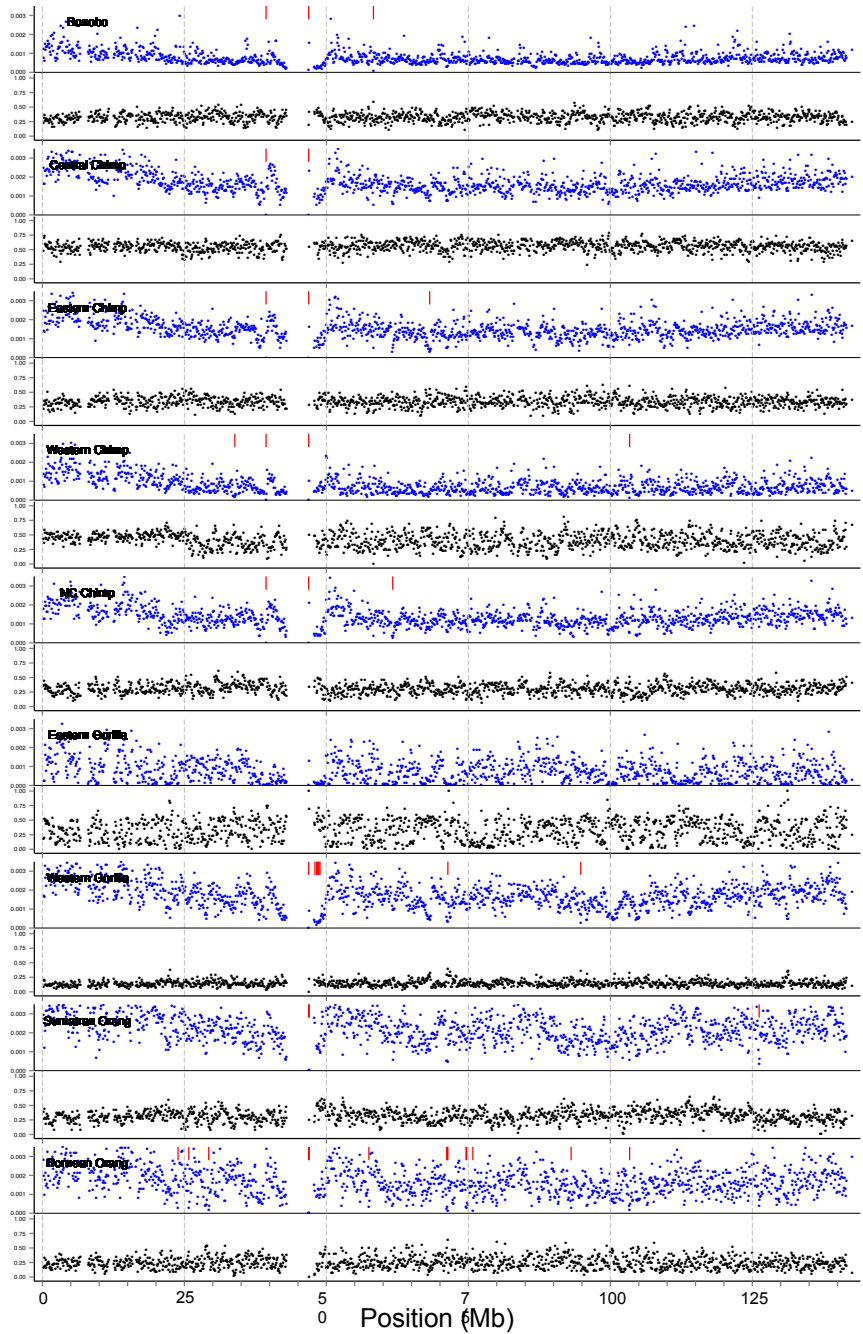

Supplementary Figure 3b. **The pattern of diversity and site-frequency spectrum in autosome.** The pi value (blue) and the proportion of singletons (black) along chromosome 8, which has a comparable size with X chromosomes. Putative candidate sweeps are identified from the lower pi values than 20% of mean of the chromosomes and denoted as red bars.

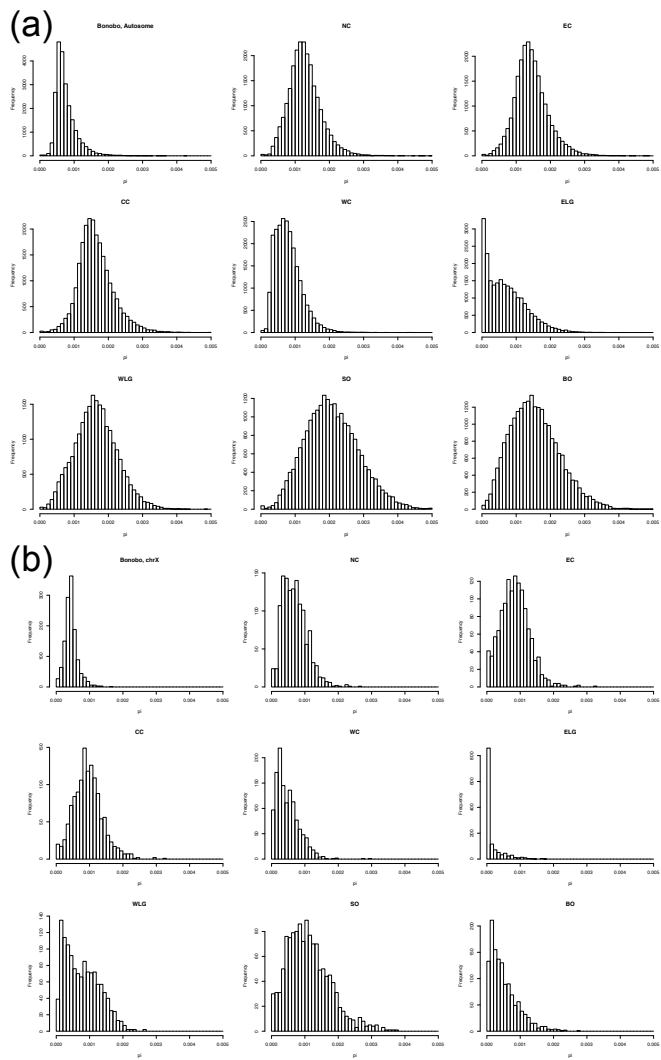

Supplementary Figure 4. **The distribution of pi values of X chromosomes and autosomes** The histrogram showing the distribution of pi values of (a) autosomes and (b) X chromosomes. Abbreviations: B: Bonobo, CC: Central Chimpanzee, EC: Eastern Chimpanzee, WC: Western Chimpanzee, NC: Nigeria-Cameron Chimpanzee , ELG: Eastern Lowland Gorilla, WLG: Western Lowland Gorilla, SO: Sumatran Orang-utan, BO: Bornean Orang-utan

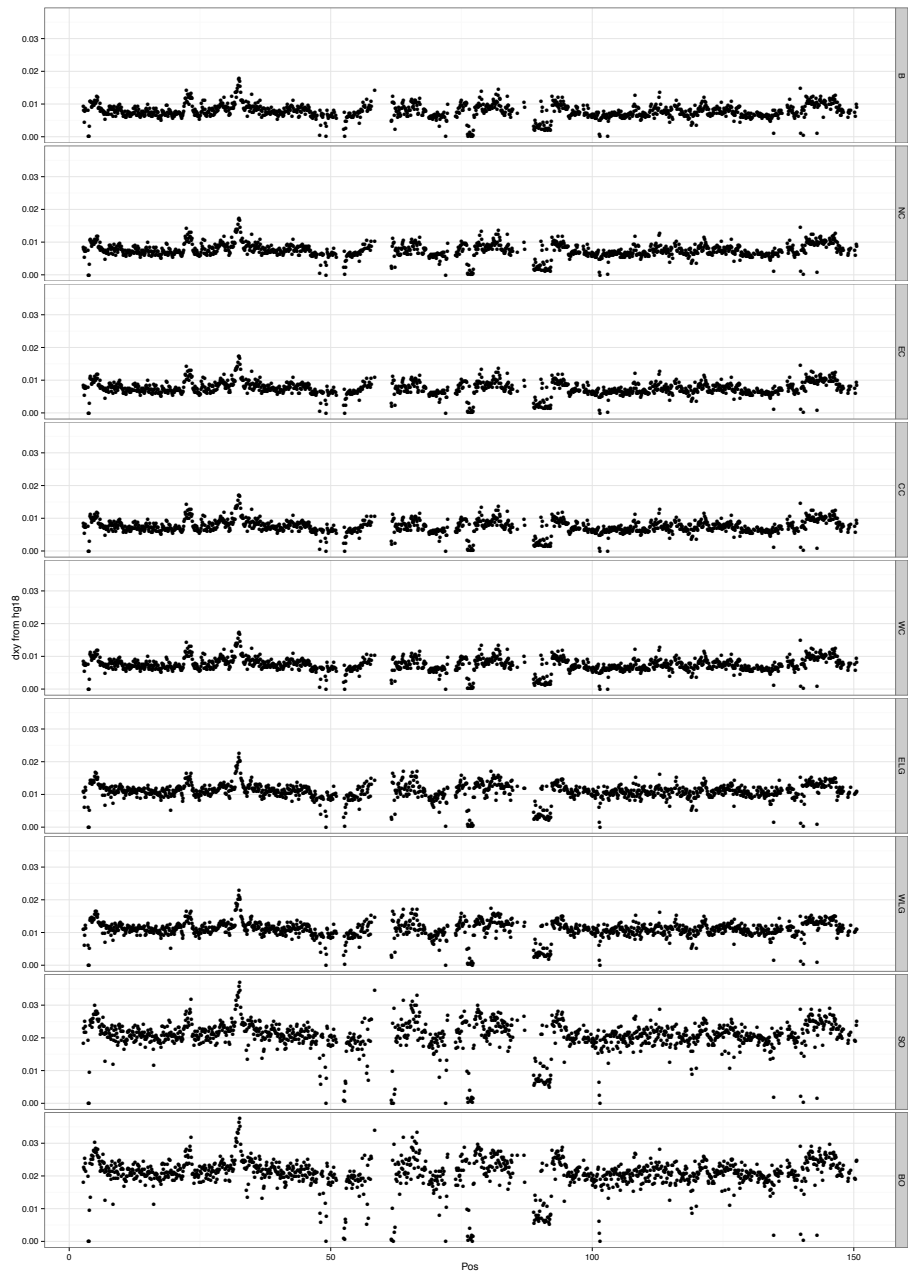

Supplementary Figure 5. The divergence pattern along chromosome X. The Dxy values calculated by the sequence comparison with hg18. (Abbreviations: B: Bonobo, CC: Central Chimpanzee, EC: Eastern Chimpanzee, WC: Western Chimpanzee, NC: Nigeria-Cameron Chimpanzee , ELG: Eastern Lowland Gorilla, WLG: Western Lowland Gorilla, SO: Sumatran Orang-utan, BO: Bornean Orang-utan).

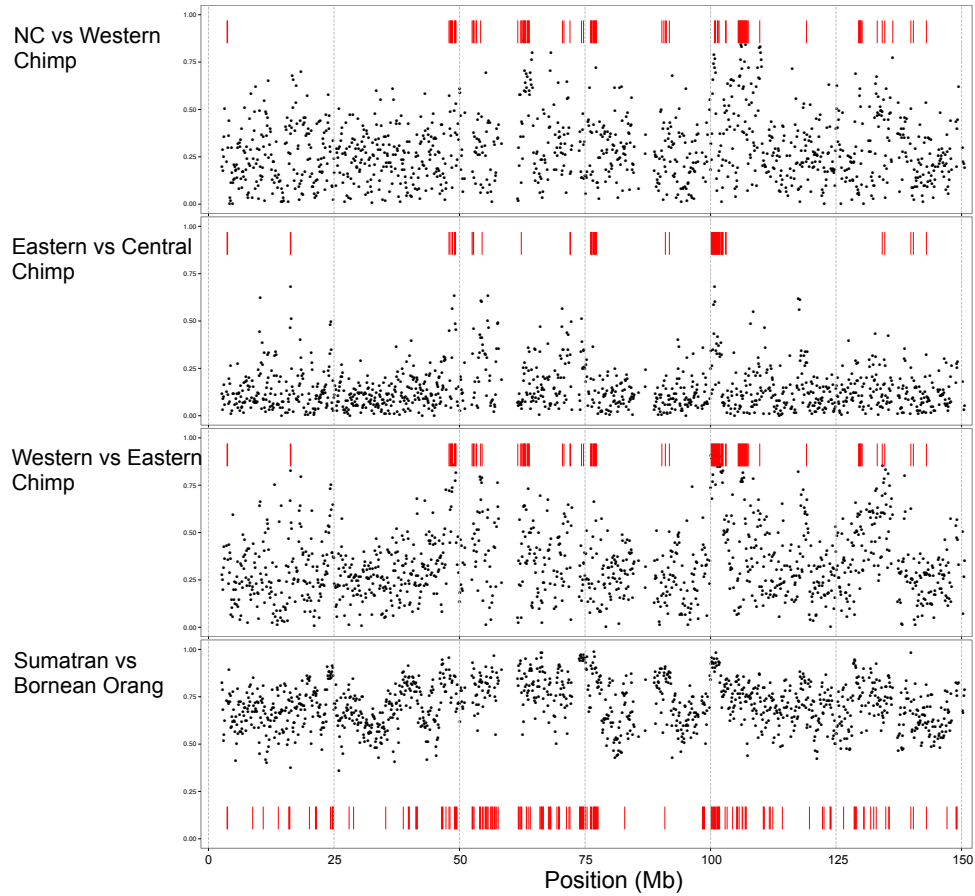

Supplementary Figure 6. **The Fst values along chromosome X**. The Fst values (black dots) are calculated from the pairs of the NC chimpanzee and Western chimpanzee, Eastern chimpanzee and Central chimpanzee, Western chimpanzee and Eastern chimpanzee, and Sumatran orang-utan and Bornean orang-utan. The red bars denote the putative sweeps identified from the lower diversity than 20% of mean pi values of the species.

24